\documentclass[11pt]{article}

\usepackage{graphicx} 
\usepackage{hyperref}
\usepackage{graphicx}
\usepackage{amssymb}
\usepackage{amsmath}
\usepackage{color}
\usepackage{url}
\usepackage{orcidlink}

\title{\large\bf GUP-induced Gravitational Baryogenesis in the Primordial Universe}

\author{\normalsize\sc Surendra Kumar Gour\thanks{\tt s.gour@iitg.ac.in} \ and Malay K. Nandy\,\orcidlink{0000-0002-2640-0497}\thanks{\tt mknandy@iitg.ac.in \rm (Corresponding Author)}\\ 
 {\small\em Department of Physics, Indian Institute of Technology Guwahati, Guwahati 781 039, India}}
 
\date{\small (26 November 2026)} 

\begin{document}

\maketitle

\begin{abstract}
We investigate gravitational baryogenesis within the quantum gravity inspired framework of Generalized Uncertainty Principle (GUP). Since the GUP-induced phase-space measure significantly alters the thermodynamic properties of the primordial fluid, this scenario gives rise to an equation of state favorable for realizing gravitational baryogenesis. Within the GUP framework, we derive the energy density and pressure of the cosmic fluid composed of Standard Model bosons and fermions. This leads to a strong departure from the standard radiation-dominated equation of state, leading to the equation-of-state parameter $w\neq \frac{1}{3}$. Such departure generates a nonvanishing baryon chemical potential and hence a non-zero net baryon number density via gravitational baryogenesis. Moreover, the GUP-induced photon and net baryon number densities, $n_{\gamma}$ and $n_B$, exhibit strong departures from their standard thermodynamic counterparts. Comparing the resulting baryon-to-photon ratio, $\eta= \frac{n_B}{n_{\gamma}}$, against the observational value  constrains the GUP parameter. This scenario provides a connection of quantum-gravity-motivated modification to statistical thermodynamics of the cosmic fluid with the origin of matter-antimatter asymmetry in the early universe.\\

\noindent
Keywords: Baryon/Lepton Number Violation; CP Violation; Baryo-and Leptogenesis; Models of Quantum Gravity

\end{abstract}

\pagebreak

\tableofcontents


\section{Introduction}

According to the Standard Model (SM) of particle physics, baryons and antibaryons can only be produced in equal numbers in the early universe so that they would be completely annihilated into radiation when the universe cooled leaving a completely matter-less universe. However, the observed existence of baryonic matter demands some mechanism that would generate more baryons over antibaryons. 

The observed baryonic asymmetry is one of the fundamental unresolved problems in the standard cosmology and particle physics, characterized by the baryon-to-photon ratio, $\eta = \frac{n_B}{n_\gamma}$, where $n_B=n_b -n_{\bar b}$ is the net baryon number density, the difference between the baryonic and antibaryonic contributions, $n_b$ and $n_{\bar b}$, and
$n_{\gamma}$ is the photon number density the early universe. 

CMB WMAP observations and BBN predictions \cite{Burles-PRD-2001, Burles-2001, Bennett-2003} indicate that the baryon-to-photon ratio has a mean value of $6.1 \times 10^{-10}$, which is also determined from Planck 2018 data \cite{Planck-2018, Planck-2024, PDG-2024}. Although this asymmetry is very small, explaining its origin is an important question in cosmology. 

In its theoretical formulation, Sakharov \cite{Sakharov-1967} conjectured three conditions for dynamically generating baryon asymmetry, namely, baryon-number violation, violation of charge ($C$) conjugation and charge-parity ($CP$) symmetry, and departure from thermal equilibrium. These conditions formed the basis for several proposed baryogenesis scenarios.

Some of the proposed mechanisms to explain baryon asymmetry include baryogenesis in grand unified theory \cite{Georgi-Glashow-1974,  Dimopoulos-1978, Yoshimura-1978-A, Yoshimura-1978-B, Weinberg-1979, Kolb-1980, Kolb-1990-book, Weinberg-book}, electroweak baryogenesis \cite{Kuzmin-1985, Shaposhnikov-1986, Cline-2006}, Affleck-Dine mechanism \cite{Affleck-Dine-1985, Allahverdi-2012}, leptogenesis \cite{Fukugita-1986}, spontaneous baryogenesis \cite{Cohen-1987, Cohen-1988, Cohen-1991}, and gravitational baryogenesis \cite{Davoudiasl-2004}. 

Among these possibilities, gravitational baryogenesis provides an interesting framework in which the evolution of the gravitational field in the presence of its coupling with the net baryon current can influence the generation of matter-antimatter asymmetry.  In a time-dependent cosmological background, this interaction can induce an effective chemical potential that distinguishes baryons from antibaryons. 

Following the development of  gravitational baryogenesis mechanism, subsequent studies explored its implementation within frameworks of modified gravity theories \cite{Hong-2004, Lambiase-2006, Pizza-2015, Ramos-2017, Whittingham-2026}. This extension provided a new direction for investigating the generation of matter–antimatter asymmetry.  This allowed for examining the viability of gravitational baryogenesis in a broader cosmological context.

However, a complete understanding of the earliest cosmological epochs requires a consistent treatment of quantum-gravitational effects. The Generalized Uncertainty Principle (GUP) provides a phenomenological framework for incorporating quantum-gravity effects by introducing a minimum measurable length into the system \cite{Achim-1995, Pikovski-2012, Shreya-2018, Fridman-2023}. In this context, a GUP-corrected gravitational baryogenesis mechanism based on the holographic principle has been proposed as a possible explanation for the observed baryon asymmetry of the universe \cite{Das-2021, Feng-2022, Luo-2023}.

In this paper, we adopt an alternative approach to GUP-assisted gravitational baryogenesis by investigating the thermodynamic properties of the cosmic fluid. In the GUP framework, the phase-space measure undergoes a substantial change leading to significant deviation in the thermodynamic behaviour of the cosmic fluid from that predicted by standard thermodynamics. This modification leads to substantial change in the resulting in an equation-of-state parameter $w\neq \frac{1}{3}$, which also depends on the GUP parameter and the temperature of the fluid.  We find the for appropriate choices of the model parameters and the temperature, the gravitational baryogenesis mechanism can successfully account for the observed baryon-to-photon ratio of the universe.

The remainder of the paper is organized as follows. In this section, we next provide an overview of the Generalized Uncertainty Principle (GUP) and also discuss the basic framework of gravitational baryogenesis. Section~\ref{sec-baryo} presents the thermodynamic derivation of  baryonic number density in the GUP framework. Section \ref{sec-photon} derives the GUP-modified photon number density. In Sections \ref{sec-brt} and \ref{sec-frt}, we obtain the energy density and pressure of the  bosonic and fermionic components of radiation in the modified thermodynamic setup of the GUP framework. Section \ref{sec-BA} explicitly obtains the expression for baron-to-photon ratio in the framework of gravitational baryogenesis employing the GUP-modified thermodynamic quantities derived in the earlier sections. Finally, Section \ref{sec-disc} concludes the paper with a discussion.

The detailed mathematical evaluations of the integrals involved in calculating the baryon number density, photon number density, bosonic and fermionic energy density and pressure are presented in Appendices (\ref{sec-A}), (\ref{sec-B}), (\ref{sec-C}), (\ref{sec-E}), (\ref{sec-D}) and (\ref{sec-F}), respectively.

\subsection{Generalized Uncertainty Principle}
\label{subsec-gup}

Various formulations of quantum gravity suggests that the geometry of the space-time has a grainy or foamy structure perceptible at ultra high energy scales. This situation can be incorporated by including a minumum length scale, by postulating a Generalized Uncertainty Principle (GUP) of the form \cite{Achim-1995, Pikovski-2012, Shreya-2018, Fridman-2023}
\begin{equation}
\Delta x\,\Delta p \geq\frac{\hbar}{2}\left[1+\beta(\Delta p)^2\right],
\end{equation}
where $\beta \sim \frac{\ell^2}{\hbar^2}$ introduces the length scale $\ell$ into the uncertainty relation.

Such a formulations can be represented through the modified commutation relation, with $\hbar=1$,
\begin{equation}
    \left[\hat{x}_i, \hat{p}_j\right]= i\,\left[\delta_{ij}(1+\beta\, \hat{\bf p}^2) + 2\beta p_i p_j \right],
\end{equation}
where  $\beta= \beta_0/M_P^2$, with the GUP parameter $\beta_0$ remaining undetermined. However, different  values for its upper bounds have been estimated by considering various simple quantum mechanical systems as well as gravitational systems. In fact, the estimated upper bounds of $\beta_0$  varies from  $10^{18}$ to $10^{90}$ depending on the physical system under consideration \cite{Das-2008, Bawaj-2015, Neves-2019, Lambiase-2018, Okcu-2021}.

Within the GUP framework, phase-volume conservation under infinitesimal time translation requires a modification in the invariant phase-space measure \cite{Chang-2002}, given by 
\begin{equation}
d^3x d^3p \longrightarrow \frac{d^3xd^3p}{(1+\beta p^2)^3}.
\label{eq-measure}
\end{equation}

This change in the phase-space measure has a profound effect on the thermodynamic behaviour of any ultra-relativistic fluid.

\subsection{Gravitational Baryogenesis }
\label{subsec-gb}

Gravitational baryogenesis \cite{Davoudiasl-2004} is a mechanism for baryon asymmetry that is generated via coupling between the time-dependent gravitational background and the baryon current.  The gravitational baryogenesis interaction, which is CP-violating, is taken to be
\begin{equation}
\mathcal{S}_{\rm int}=\frac{1}{M_*^2}\int d^4x\,\sqrt{-g}\,  J_B^\mu (\partial_\mu R) \equiv \int d^4x\,\sqrt{-g}\, \mathcal{L}_{\rm int}  ,
\end{equation}
where  $R$ is the Ricci scalar, $J_B^\mu=(n_B,\mathbf{J}_B)$ is the net baryon four-current, and $M_*$ is the cutoff scale which is associated with the effective interaction.

The baryon current has the form $J_{b}^\mu=q_b \bar{\psi} \gamma^{\mu}\psi$, where $q_b=+1$ ( $q_{\bar b}=-1$) is charge associated with the baryon (antibaryons) current.

Since we assume the universe to be homogeneous and isotropic on large scales, the Ricci scalar depends on time alone so that  
\begin{equation}
\mathcal{L}_{\rm int}=\frac{1}{M_*^2}J_B^\mu \partial_\mu R= \frac{1}{M_*^2}J_B^0\partial_0 R= \frac{1}{M_*^2}n_B \dot{R}.
\label{eq-Lint}
\end{equation}

The Gibbs free energy density of the baryonic fluid is given by
\begin{equation}
G= \mu_b n_b+\mu_{\bar b} n_{\bar b}= \mu_b n_B,
\label{eq-gibbs}
\end{equation}
where $\mu_b$ and $\mu_{\bar b}=-\mu_b$ are the baryonic and antibaryonic chemical potentials, respectively, whereas $n_B=n_b -n_{\bar b}$ is the net baryon number density, with $n_b$ and $n_{\bar b}$ the baryonic and antibaryonic number densities, respectively.

Equating (\ref{eq-Lint}) and (\ref{eq-gibbs}), the baryonic chemical potential is identified as 
\begin{equation}
\mu_b=\frac{\dot R}{M_*^2}.
\label{eq-mubrdot}
\end{equation}

We consider a spatially flat Friedmann-Lemaître-Robertson-Walker (FLRW) universe \cite{Friedmann-1922, Lema-1927, Robertson-1936, Walker-1937, Barrow-2012} described by $ds^2=-dt^2+S^2(t)d\mathbf{x}^2$,
where $S(t)$ is the scale factor. 

The Einstein-Hilbert action 
\begin{equation}
\mathcal{S}_{\rm EH}= \frac{1}{2} M_P^2 \int d^4x \sqrt{-g} R + \mathcal{S}_m,
\end{equation}
with $M_P^2=\frac{1}{8\pi G}$ the reduced Planck mass, leads to the Einstein Field equations 
\begin{equation}
R_{\mu \nu}-\frac{1}{2} g_{\mu \nu} R = \frac{1}{M_P^2} T_{\mu \nu},
\end{equation}
where the energy-momentum tensor for a perfect fluid has the form 
\begin{equation}
T_{\mu \nu}= (\rho +P)u_{\mu}u_{\nu}+P g_{\mu \nu},
\end{equation}
with $u^{\mu}$ the four-velocity, $\rho$ and $P$ are the energy density and pressure of the fluid in the comoving frame.

The trace of the Einstein field equation turns out to be 
\begin{equation}
R= \frac{1}{M_P^2} \left(\rho -3P\right)=\frac{1}{M_P^2}(1-3w)\rho,
\label{eq-trace}
\end{equation}
with the equation of state $P=w\rho$. Thus
\begin{equation}
\dot{R}= \frac{1}{M_P^2}(1-3w)\dot{\rho}.
\label{eq-rdot}
\end{equation}

The $00$ component of Einstein field equation gives the Friedmann equation, 
\begin{equation}
H^2= \frac{1}{3M_P^2} \rho,
\label{eq-Fried}
\end{equation}
where $H=\frac{\dot{S}}{S}$ is Hubble expansion parameter.

Using (\ref{eq-Fried}) and $P=w \rho$, the equation of continuity $\dot{\rho}+ 3H(\rho+P)=0$ yields 
\begin{equation}
\dot{\rho}=- \frac{\sqrt{3\rho}}{M_P}(1+w)\rho.
\label{eq-rhodot}
\end{equation}
Substituting equation (\ref{eq-rhodot}) in (\ref{eq-rdot}), one obtains
\begin{equation}
\dot{R}= -\frac{\sqrt{3}}{M_P^3}(1-3w)(1+w)\rho^{3/2}
\label{eq-rdot1}
\end{equation}

Using the relation (\ref{eq-mubrdot}), the baryonic chemical potential is obtained as
\begin{equation}
\mu_b = \frac{\dot{R}}{M_{*}^2}= -\frac{\sqrt{3}}{M_{*}^2M_P^3}(1-3w)(1+w)\rho^{3/2},
\label{eq-mubaryon}
\end{equation}
which is expressed in terms of the energy density $\rho$ and the equation of state parameter $w=\frac{P}{\rho}$.

In the conventional radiation-dominated universe, equation (\ref{eq-mubaryon}) vanishes identically due to the equation-of-state parameter taking the value $w=1/3$. Consequently, since the net baryon number density is proportional to the baryon chemical potential, $n_B\propto\mu_b$, the gravitational baryogenesis mechanism fails to generate any baryon asymmetry during the radiation-dominated era.

Nevertheless, baryogenesis may take place in the early universe, particularly during the reheating epoch following inflation. In this context, it is essential to investigate the possible modifications to the thermodynamic properties of the primordial universe arising from the existence of a minimal length scale, as naturally incorporated within the framework of the Generalized Uncertainty Principle (GUP) \cite{Achim-1995}.

In this paper, we demonstrate that the thermodynamic properties of the radiation-dominated universe undergo significant modifications in the GUP framework, leading to an equation-of-state parameter $w\neq 1/3$. As a consequence, the net baryon number density $n_B$ acquires a finite, non-vanishing value, thereby facilitating the generation of a baryon asymmetry in the early universe through the mechanism of gravitational baryogenesis.

\section{Baryonic Number Density in GUP }
\label{sec-baryo}

For a fermionic gas consisting of baryons and antibaryons in thermal equilibrium at temperature $T$, the Fermi–Dirac distributions are given by
\begin{equation}
f_b(E)=\frac{1}{\exp\left[\frac{(E-\mu_b)}{T}\right]+1} \quad \text{and} \quad f_{\bar b}(E)=\frac{1}{\exp\left[\frac{(E+\mu_b)}{T}\right]+1},
\end{equation}
where $+\mu_b$ and $-\mu_b$ are the chemical potentials of the baryon and antibaryon, respectively. 

In the GUP scenario, the phase-space measure alters in accordance with equation (\ref{eq-measure}), giving the baryon and antibaryon number densities $n_b$ and $n_{\bar b}$ as 
\begin{equation}
n_{b,\bar b} =\frac{g_b}{2\pi^2}\int_0^\infty\frac{p^2\,dp}{(1+\beta p^2)^3}\frac{1}{\exp\left[(E \mp \mu_b)/T\right]+1},
\end{equation}
where $g_b$ represents the number of baryonic degrees of freedom (which equals the number of antibaryonic degrees of freedom $g_{\bar b}$), and $E=\sqrt{p^2+m_b^2} \approx p$ for an ultra-relativistic gas in which case the baryon mass $m_b$ can be neglected.

Consequently, we obtain the net baryon number density from 
\begin{equation} 
n_B=n_b-n_{\bar b}= \frac{g_b}{2\pi^2}\int_0^\infty  \frac{p^2 dp}{(1+\beta p^2)^3}\left[\frac{1}{e^{(p-\mu_b)/T}+1}-\frac{1}{e^{(p+\mu_b)/T}+1}\right].
\end{equation}

The integral involves the GUP measure $(1+\beta p^2)^{-3}$. Since the GUP parameter $\beta M_P^2$ is usually a large number \cite{Das-2008}, an expansion about $\beta=0$ is not physically justifiable. Consequently, the integral must be evaluated generally for any $\beta$.

Introducing the dimensionless variables $x=\frac{p}{T}$ and $\tilde{\mu}=\frac{\mu_b}{T}$ and writing $\frac{1}{\beta T^2}=a^2$, we have\footnote{In our notation, $a=\frac{1}{T \sqrt{\beta}}$ is not the scale factor. Instead, we have used the symbol $S$ for the scale factor in the line element $ds^2=-dt^2+S^2(t)d\mathbf{x}^2$, so that the Hubble rate is defined through $H=\frac{\dot{S}}{S}$.}
\begin{equation}
n_B=\frac{g_b T^3}{2\pi^2} \left(\frac{1}{\beta^3 T^6}\right) \int_0^\infty \frac{x^2 dx}{\left(a^2+x^2\right)^3}\left[\frac{1}{e^{x-\tilde{\mu}}+1}-\frac{1}{e^{x+\tilde{\mu}}+1}\right].
\label{eq-n_B}
\end{equation}

Writing $\beta M_P^2= \beta_0$, we have $a^2=\frac{M_P^2}{\beta_0 T^2}$. Since the upper bound for $\beta_0$ is usually large, we expect that the dimensionless quantity $a$ would be a small number.

It is however important to note that we cannot simply neglect $a$ despite its smallness since substituting $a=0$ would lead to an infrared divergence in the integral (\ref{eq-n_B}). Consequently, we must evaluate this integral for any small value of the parameter $a$.  

Expanding the integrand with $\tilde{\mu}\ll1$, we have
\begin{equation}
\frac{1}{e^{x-\tilde{\mu}}+1} - \frac{1}{e^{x+\tilde{\mu}}+1} \simeq 2\tilde{\mu} \frac{e^x}{(e^x+1)^2},
\end{equation}
to first order in $\tilde{\mu}$, so that equation (\ref{eq-n_B}) reduces to
\begin{equation}
n_B=\frac{g_b T^2 \mu_b}{\pi^2} a^{6}  \int_0^\infty dx\,\frac{x^2}{\left(a^2+x^2\right)^3}\frac{e^x}{(e^x+1)^2}.
\label{eq-netbaryon}
\end{equation}

Thus, we have to evaluate the integral 
\begin{equation}
I_1(a)=\int_0^\infty \frac{x^2 dx}{\left(a^2+x^2\right)^3}\frac{e^x}{(e^x+1)^2},
\label{eq-I1}
\end{equation}
for any general value of the small parameter $a$. 

Since the result of the integration $I_1(a)$ must depend on $a$, its evaluation has to be carried out with care. Consequently, we evaluate the integral (\ref{eq-I1}) employing the convolution theorem of Mellin transforms as detailed in Appendix (\ref{sec-A}). The result is 

\begin{equation}
  I_1(a)=\frac{\pi}{64a^3}-\frac{3\pi}{256a}+\frac{31\zeta(5)}{8\pi^4}+\cdots.
\end{equation}

Using this result, the net baryon number density given by equation (\ref{eq-netbaryon}) takes the form
\begin{equation}
n_B=  \frac{g_b  T^2 \mu_b}{64\pi}  a^3 \left[ 1 - \frac{3}{4} a^2 + \frac{248\zeta(5)}{\pi^5} a^3 +\cdots \right].
\label{eq-finalbaryon} 
\end{equation}

\section{Photon Number Density in GUP}
\label{sec-photon}

For a photon gas in thermal equilibrium at temperature $T$, the Bose distribution is given by
\begin{equation}
f_\gamma(p)= \frac{1}{e^{p/T}-1}.
\end{equation}

Since the phase-space measure alters in accordance with equation (\ref{eq-measure}) in the GUP scenario, we obtain the photon number density as
\begin{align}
n_\gamma=\frac{1}{\pi^2}\int_0^\infty\frac{p^2\,dp}{(1+\beta p^2)^3} \frac{1}{(e^{p/T}-1)}.
\end{align}

Introducing the dimensionless variable $x=\frac{p}{T}$, and writing $\frac{1}{\beta T^2}=a^2$, the photon number density can be written as
\begin{equation}
    n_\gamma = \frac{T^3}{\pi^2} a^6 \int_0^\infty \frac{1}{(a^2+x^2)^3} \frac{x^2dx}{e^x-1}.
\label{eq-photon}    
\end{equation}

Consequently, we have to evaluate the integral 
\begin{equation}
    I_2(a) = \int_0^\infty \frac{x^2}{(a^2+x^2)^3} \frac{dx}{e^x-1}.
\label{eq-I2}    
\end{equation}

As explained earlier, this integral must be evaluated without expanding about $a=0$. Consequently, we evaluate the integral (\ref{eq-I2}) employing the convolution theorem of Mellin transforms as detailed in Appendix (\ref{sec-B}). The result is 
\begin{equation}
I_2(a)= \frac{1}{4a^4}-\frac{\pi}{32a^3}+\frac{1}{48a^2} + \cdots.
\end{equation}

Using this result, the photon number density given by equation (\ref{eq-photon}) assumes the form
\begin{equation}
n_\gamma = \frac{ T^3}{4\pi^2} a^2 \left[1-\frac{\pi a}{8} +\frac{a^2}{12}+ \cdots \right].
\label{eq-finalphoton}
\end{equation}

\section{Bosonic Radiation in GUP}
\label{sec-brt}

In the mechanism of gravitational baryogenesis, it is natural to guess that the decoupling temperature would be very high, making the universe highly radiation-dominated. Consequently, all Standard Model degrees of freedom are expected to be excited. We therefore first consider the thermodynamic behaviour of the bosonic radiation degrees of freedom in this radiation-dominated epoch.

\subsection{Bosonic Radiation Energy Density}
\label{subsec-bre}

Since the phase-space measure in the GUP regime is given by equation (\ref{eq-measure}), we obtain the bosonic radiation energy density as
\begin{equation}
    \rho_\gamma = \frac{g_{\gamma}}{2\pi^2}\int_0^\infty \frac{p^3dp}{(1+\beta p^2)^3} \frac{1}{e^{p/T}-1},
\label{eq-xy1}    
\end{equation}
where $g_{\gamma}$ represents the number of all bosonic ultra-relativistic degrees of freedom.

Introducing the dimensionless variable $x=\frac{p}{T}$, and writing $\frac{1}{\beta T^2}=a^2$, the bosonic radiation energy density (\ref{eq-xy1}) can be written as
\begin{equation}
    \rho_\gamma = \frac{g_{\gamma} T^4}{2\pi^2} a^6 \int_0^\infty \frac{1}{(a^2+x^2)^3} \frac{x^3dx}{e^x-1}.
\label{eq-photone}    
\end{equation}

Consequently, we have to evaluate the integral 
\begin{equation}
    I_3(a) = \int_0^\infty \frac{1}{(a^2+x^2)^3} \frac{x^3dx}{e^x-1}.
\label{eq-I3}    
\end{equation}

As explained earlier, this integral must be evaluated without expanding about $a=0$. As shown in  Appendix (\ref{sec-C}), this integral can be evaluated by considering an auxiliary integral in addition to employing Binet's second representation of the digamma function. The result is 
\begin{equation}
I_3(a)= \frac{\pi}{16a^3}-\frac{1}{8a^2}+\frac{3\zeta(2)}{32 \pi a}+ \cdots.
\end{equation}

Using this result, the bosonic radiation energy density given by equation (\ref{eq-photone}) assumes the form
\begin{equation}
\rho_\gamma = \frac{g_{\gamma}T^4}{32\pi}a^3\Bigg[ 1-  \frac{2a}{\pi}+ \frac{3\zeta(2)}{2\pi^2}
a^2 + \cdots \Bigg].
\label{eq-rhogam2}
\end{equation}

\subsection{Bosonic Radiation Pressure}
\label{subsec-brp}

To obtain the pressure of the bosonic radiation, we employ the grand partition function \cite{Landau-1980} 
\begin{equation}
\mathcal{Z}_{\gamma}=\prod_{\mathbf{p}}\left\{\sum_{n_p}\exp\left[-\left(\frac{E-\mu_{\gamma}}{ T}\right)n_p\right]\right\}^{g_{\gamma}}.
\end{equation}

Since $n_p=0,1, 2, \cdots, \infty $ for bosons, the grand potential for the bosonic radiation, $\Omega_{\gamma}=-T\ln\mathcal{Z}_{\gamma}$ turns out to be
\begin{equation}
\Omega_{\gamma}=g_{\gamma}T\sum_{\mathbf{p}}\ln\left\{1-\exp\left(-\frac{E}{T}\right)\right\},
\end{equation}
with $\mu_{\gamma}=0$.

Noting the relation $\Omega_{\gamma}=-P_{\gamma}V$, using the phase-space measure (\ref{eq-measure}) in the GUP regime, we obtain the  radiation pressure as 
\begin{equation}
 P_{\gamma}= -\frac{g_{\gamma}T}{2\pi^2} \int_0^\infty\frac{ p^2\,dp}{(1+\beta p^2)^3} \ln \left\{ 1-\exp\left(-\frac{p}{T}\right)\right\},
\end{equation}
with $E=\sqrt{p^2+m^2} \approx p$ for an ultra-relativistic bose gas.

With the dimensionless variable $x=\frac{p}{T}$, and writing $\frac{1}{\beta T^2}=a^2$, we thus have
\begin{equation}
 P_{\gamma}= -\frac{g_{\gamma}T^4}{2\pi^2} a^6 \int_0^\infty \frac{x^2 dx}{(a^2+x^2)^3} \ln\left(1-e^{-x}\right).
\label{eq-Pgam}
\end{equation}

Consequently, we have to evaluate the integral 
\begin{equation}
I_4(a)=\int_0^\infty \frac{x^2 dx}{(a^2+x^2)^3} \ln\left(1-e^{-x}\right).
\label{eq-I6}
\end{equation}

We evaluate this integral (\ref{eq-I6}) without expanding about $a=0$ by employing the convolution theorem of Mellin transforms as detailed in Appendix (\ref{sec-E}). The result is 
\begin{align}
  I_4(a)=& \frac{\pi \ln a}{16a^3}-\frac{1}{8a^2}+\frac{\pi}{128 a} + \cdots.
\label{eq-I6F}  
\end{align}

Substituting (\ref{eq-I6F}) in (\ref{eq-Pgam}), we have 
\begin{equation}
 P_{\gamma}= -\frac{g_{\gamma}T^4}{32\pi} a^3 \left[ \ln a-\frac{2 a}{\pi}+\frac{a^2}{8} + \cdots \right].
\label{eq-Pgam2}
\end{equation}

\section{Fermionic Radiation in GUP}
\label{sec-frt}

In the radiation-dominated epoch, the radiation sector is also populated by all Standard Model fermionic degrees of freedom. Consequently, we also include the thermodynamic properties of the fermionic radiation degrees of freedom.

\subsection{Fermionic Radiation Energy Density}
\label{subsec-fre}

Accounting for the phase-space measure in GUP given by equation (\ref{eq-measure}), we have the fermionic radiation energy density 
\begin{equation}
  \rho_f = \frac{g_{f}}{2\pi^2}\int_0^\infty \frac{p^3dp}{(1+\beta p^2)^3} \frac{1}{e^{(p-\mu_f)/T}+1},
\label{eq-tf}    
\end{equation}
where $g_{f}$ is the number of  fermionic degrees of freedom and $\mu_f$ is the fermionic chemical potential.

With the dimensionless variable $x=\frac{p}{T}$, and writing $\frac{1}{\beta T^2}=a^2$, equation (\ref{eq-tf}) reduces to
\begin{equation}
    \rho_f = \frac{g_{f} T^4}{2\pi^2} a^6 \int_0^\infty \frac{x^3dx}{(a^2+x^2)^3} \frac{1}{e^x+1},
\label{eq-rhof0}    
\end{equation}
with $\mu_f \ll T$.

Consequently, we evaluate the integral 
\begin{equation}
    I_5(a) = \int_0^\infty \frac{x^3dx}{(a^2+x^2)^3} \frac{1}{e^x+1},
\label{eq-I4}
\end{equation}
employing the convolution theorem of Mellin transforms as detailed in Appendix (\ref{sec-D}). The result is 
\begin{equation}
I_5(a)=  \frac{1}{8 a^2} -\frac{3 \pi}{64a} + \frac{7 \zeta(3)}{8\pi^2}+ \cdots.
\label{eq-I4f}
\end{equation}

Substituting (\ref{eq-I4f}) in (\ref{eq-rhof0}), we obtain the total fermionic  energy density as
\begin{equation}
    \rho_f = \frac{g_{f} T^4}{16\pi^2} a^4 \left[1 -\frac{3 \pi a}{8} + \frac{7 \zeta(3) a^2}{\pi^2} + \cdots \right ].
\label{eq-rhof}    
\end{equation}

\subsection{Fermionic Radiation Pressure}
\label{subsec-frp}

To obtain the pressure of the fermionic sector of radiation, we use the grand partition function, expressed by \cite{Landau-1980} 
\begin{equation}
\mathcal{Z}_{f}=\prod_{\mathbf{p}}\left\{\sum_{n_p}\exp\left[-\left(\frac{E-\mu_{f}}{ T}\right)n_p\right]\right\}^{g_{f}}.
\end{equation}
 
Since $n_p=0,1$ for fermions, the grand potential for the fermionic radiation sector, $\Omega_f=-T\ln\mathcal{Z}_f$, is given by
\begin{equation}
\Omega_f=-g_fT\sum_{\mathbf{p}}\ln\left\{1+\exp\left(-\frac{E}{T}\right)\right\},
\end{equation}
noting that $\mu_f \ll T$.

Using the relation $\Omega_f=-P_fV$, and the phase-space measure (\ref{eq-measure}) in the GUP regime, we obtain the  fermionic radiation pressure as 
\begin{equation}
 P_f= \frac{g_fT}{2\pi^2} \int_0^\infty\frac{ p^2\,dp}{(1+\beta p^2)^3} \ln \left\{ 1+\exp\left(-\frac{p}{T}\right)\right\},
\end{equation}
with $E=\sqrt{p^2+m^2} \approx p$ for an ultra-relativistic fermi gas.

Employing the dimensionless variable $x=\frac{p}{T}$, and using $\frac{1}{\beta T^2}=a^2$, we obtain
\begin{equation}
 P_f= \frac{g_fT^4}{2\pi^2} a^6 \int_0^\infty \frac{x^2 dx}{(a^2+x^2)^3} \ln\left(1+e^{-x}\right).
 \label{eq-Pb}
\end{equation}

Consequently, we have to evaluate the integral 
\begin{equation}
I_6(a)=\int_0^\infty \frac{x^2 dx}{(a^2+x^2)^3} \ln\left(1+e^{-x}\right),
\label{eq-I5}
\end{equation}
by the convolution theorem of Mellin transforms as detailed in Appendix (\ref{sec-F}). The result is 
\begin{align}
  I_6(a)= \frac{\pi \ln 2}{16a^3}-\frac{1}{8a^2}+\frac{3\pi}{128 a}  + \cdots.
\label{eq-I5F}  
\end{align}

Substituting (\ref{eq-I5F}) in (\ref{eq-Pb}), we thus obtain the fermionic radiation pressure as
\begin{equation}
 P_f= \frac{g_fT^4}{32\pi} a^3  \left[ \ln 2-\frac{2 a}{\pi}+\frac{3 a^2}{8} + \cdots \right].
 \label{eq-Pb2}
\end{equation}

\section{Baryon Asymmetry in GUP }
\label{sec-BA}

Using the expression for baryonic chemical potential given by (\ref{eq-mubaryon}) in gravitational baryogenesis, we obtain the expression for the net baryon density from (\ref{eq-finalbaryon}) in the GUP regime as
\begin{equation}
n_B=  -\frac{\sqrt{3} g_b  }{64\pi} \frac{T^2a^3}{M_{*}^2M_P^3}(1-3w)(1+w)\rho^{3/2} \left[ 1 - \frac{3}{4} a^2 + \frac{248\zeta(5)}{\pi^5} a^3 +\cdots \right].
\label{eq-finalbaryon2} 
\end{equation}

We recall that the energy density $\rho$ and the equation of state parameter $w$ in this equation (\ref{eq-finalbaryon2}) can be traced back to the Friedmann equation (\ref{eq-Fried}) and the trace of the Einstein field equation (\ref{eq-trace}). Consequently, these quantities must be identified with the total energy density and total pressure of the cosmic fluid. 

Since the decoupling temperature is expected to be very high in gravitational baryogenesis, the universe must be populated with ultra-relativistic bosonic and fermionic species of the Standard Model of particle physics. Consequently, we identify the energy density and pressure of the cosmic fluid by $\rho =\rho_{\gamma}+ \rho_{f}+ \rho_{\bar f}$ and $P =P_{\gamma}+ P_{f}+ P_{\bar f}$, which gives $w=\frac{P}{\rho}$. Consequently, the total energy density is obtained from (\ref{eq-rhogam2}) and (\ref{eq-rhof}) as 
\begin{equation}
    \rho = \frac{g_{\gamma}T^4}{32\pi}a^3\Bigg[ 1-  \frac{2a}{\pi}+ \frac{3\zeta(2)}{2\pi^2}
a^2  + \cdots \Bigg] + \frac{(g_{f}+ g_{\bar f}) T^4}{16\pi^2} a^4 \left[1 -\frac{3 \pi a}{8} + \frac{7 \zeta(3) a^2}{\pi^2}+ \cdots \right ],
\label{eq-rhofinal}    
\end{equation}
and the total pressure obtained from (\ref{eq-Pgam2}) and (\ref{eq-Pb2}) as
\begin{equation}
 P= -\frac{g_{\gamma}T^4}{32\pi} a^3 \left[ \ln a-\frac{2 a}{\pi}+\frac{a^2}{8}  + \cdots \right] + \frac{(g_{f}+ g_{\bar f})T^4}{32\pi} a^3  \left[ \ln 2-\frac{2 a}{\pi}+\frac{3 a^2}{8} \cdots \right].
\label{eq-Pfinal}
\end{equation}

Since $a$ is a small quantity, the fermionic contribution to the energy density in (\ref{eq-rhofinal}) is subdominant. On the other hand, both bosonic and fermionic contributions to the pressure in (\ref{eq-Pfinal}) are of the same order. Considering the dominant parts, we thus have  
\begin{equation}
    \rho = \frac{g_{\gamma}T^4}{32\pi}a^3,
\label{eq-rhofinal2}    
\end{equation}
\begin{equation}
 P= \frac{T^4}{32\pi} a^3 \left[- g_{\gamma} \ln a + (g_{f}+ g_{\bar f})\ln 2 \right], 
\label{eq-Pfinal2}
\end{equation}
so that the equation of state parameter, in the GUP regime, turns out to be 
\begin{equation}
w=\frac{P}{\rho}= - \ln a + \frac{(g_{f}+ g_{\bar f})}{g_{\gamma}} \ln 2.
\label{eq-w1}
\end{equation}
This implies that $w \neq \frac{1}{3}$ in the GUP regime so that the net baryon number density $n_B \neq 0$ in the GUP regime of gravitational baryogenesis described by equation \ref{eq-finalbaryon2}.

Using the fact that $a$ is a small quantity, we obtain the baryon-to-photon ratio from equations (\ref{eq-finalbaryon2}) and (\ref{eq-finalphoton}) as
\begin{equation}
\eta=\frac{n_B}{n_{\gamma}}= -\frac{\sqrt{3} \pi g_b  }{16} \frac{a}{M_{*}^2M_P^3 T}(1-3w)(1+w)\rho^{3/2}.
\label{eq-xyz}
\end{equation}
Substituting the expression for total energy density $\rho$ from (\ref{eq-rhofinal2}) in (\ref{eq-xyz}), we obtain
\begin{equation}
\eta= -\sqrt{\frac{3}{2\pi}}\frac{g_b g_{\gamma}^{3/2}}{2048} \frac{a^{11/2}T^5}{M_{*}^2M_P^3 }(1-3w)(1+w).
\label{eq-eta2}
\end{equation}

The bosonic sector of the Standard Model has a total of $g_{\gamma}=28$ degrees of freedom (photon, gluons, $W^{\pm}$, $Z^{0}$, Higgs bosons). On the other hand, the fermionic sector has $g_{f}=45$ degrees of freedom (quarks, charged leptons, neutrinos), and the corresponding anti-fermions contribute $g_{\bar f}=45$. Of these, the baryonic (quark) sector has $g_b= 36$ degrees of freedom.  

From (\ref{eq-w1}), we thus have the GUP equation of state parameter
\begin{equation}
w= -\ln \left(\frac{M_P} {T\sqrt{\beta_0}}\right) + \frac{45}{14} \ln 2,
\label{eq-w2}
\end{equation}
and from (\ref{eq-eta2}), we finally obtain the baryon-to-photon ratio
\begin{equation}
\eta= \frac{63}{64}  \sqrt{\frac{21}{2\pi}\frac{M_P^5}{ M_{*}^4T}} \frac{(3w-1)(w+1)}{\beta_0^{11/4}},
\label{eq-eta3}
\end{equation}
in the GUP regime of gravitational baryogenesis.

The observed mean value of baryon-to-photon ratio $\eta=6.1 \times 10^{-10}$, together with equations (\ref{eq-w2}) and (\ref{eq-eta3}), constrains the values of the equation of state parameter to $w=4.373$ and the GUP parameter to $\beta_0=4.328 \times 10^{4}$, for $M_*= 10^{18}$ GeV  and the decoupling temperature $T= 10^{17}$ GeV.

The enhanced value of the equation of state parameter to $w=4.373$ is a consequence of the effect of quantum gravity incorporated phenomenologically through GUP, because of which $w$ acquires dependence on $\beta_0$ and $T$ as expressed in equation (\ref{eq-w2}). Moreover, the  baryon-to-photon ratio $\eta$ acquires dependence on $\beta_0$, in addition to $M_*$ and $T$, as shown in equation (\ref{eq-eta3}).

\section{Discussion and Conclusion}
\label{sec-disc}

In this paper, we formulated a mechanism of baryogenesis within the framework of the Generalized Uncertainty Principle (GUP), incorporating the formalism of gravitational baryogenesis. The primary motivation for this investigation is to explore the possible implications of GUP-induced modifications to the thermodynamic properties of the primordial cosmic fluid and their role in generating the observed baryon asymmetry of the universe.

The Generalized Uncertainty Principle (GUP) represents a generalization of the conventional Heisenberg Uncertainty Principle (HUP), in which the existence of a fundamental length scale is incorporated into the uncertainty relation. Such a modification is motivated by the interplay between quantum mechanics and gravity, particularly in the context of quantum-gravitational effects arising during high-energy particle measurements. In this regard, the conventional Heisenberg microscope thought experiment can be extended to account for the gravitational interactions associated with the measurement process. Consequently, the conventional uncertainty relation between position and momentum undergoes a fundamental modification, thereby altering the structure of the phase space.

Given that the primordial universe was characterized by an extremely high-energy environment in which both photons and gravitons were expected to play significant roles, incorporating GUP-induced modifications into the thermodynamic description of the early universe becomes particularly well motivated. In the GUP framework, the phase-space measure is modified, potentially leading to significant deviations from conventional thermodynamic behaviour. 

These modifications can become particularly relevant in the mechanism of gravitational baryogenesis. Due to a high decoupling temperature, the relativistic degrees of freedom associated with the Standard Model of particle physics are expected to be thermally accessible.

In the mechanism of gravitational baryogenesis, the baryon asymmetry of the universe is generated through a CP-violating coupling between the four-gradient of the Ricci scalar and the net baryon four-current. This coupling introduces an effective contribution to the Gibbs free energy, giving rise to a baryon chemical potential proportional to the time derivative of the Ricci scalar, that is, $\mu_b \propto \dot{R}$.

The gravitational equations of motion, together with the continuity equation, yields the proportionality $\dot{R} \propto (1-3w)(1+w)\rho^{3/2}$, where $\rho$ denotes the total energy density of the cosmic fluid and $w$ is its equation-of-state parameter. Consequently, the baryon chemical potential $\mu_b$ becomes directly related to the thermodynamic properties of the cosmic fluid, $\rho$ and $w$. Furthermore, the quantum statistical distributions of baryons and antibaryons lead to the net baryon number density $n_B=(n_b-n_{\bar b})$, representing the difference between the baryonic and antibaryonic number densities of the cosmic fluid, determined by the proportionality $n_B\propto\mu_b$.

It is important to emphasize that, under conventional thermodynamic conditions, a radiation-dominated universe is characterized by an equation-of-state parameter $w=1/3$. Thus, $\dot{R}=0$, which consequently leads to a vanishing baryon chemical potential, $\mu_b=0$. As a result, no baryon asymmetry is generated, that is, $n_B=0$, within the conventional framework of gravitational baryogenesis.

To investigate the possibility of circumventing this limitation, we consider gravitational baryogenesis in the GUP-modified thermodynamic framework. In particular, we calculate the net baryon number density, $n_B=n_b-n_{\bar b}$, employing the modified phase-space measure prescribed by the GUP. Interestingly, we find that the net baryon number density $n_B$ retains its proportionality to the baryon chemical potential $\mu_b$ in the GUP framework, although its expression acquires additional factor dependent on the GUP parameter.

Subsequently, by considering an ultra-relativistic cosmic fluid comprising of the fermionic and bosonic sectors of the Standard Model, we derived the corresponding energy densities and pressures within GUP-modified statistical thermodynamics. Our analysis reveals that the resulting equation-of-state parameter deviates from its conventional radiation-dominated value, yielding $w\neq 1/3$. This deviation is particularly significant in the context of gravitational baryogenesis, as it leads to a nonvanishing baryon chemical potential, $\mu_b\neq 0$, and consequently generates a nonzero net baryon number density, $n_B\neq 0$. Furthermore, we find that the equation-of-state parameter $w$ exhibits an explicit dependence on the GUP parameter $\beta_0$, the temperature $T$, and the effective relativistic degrees of freedom associated with the bosonic and fermionic sectors of the Standard Model.

In order to find the baryon-to-photon ratio, $\eta=\frac{n_B}{n_\gamma}$, we derived the photon number density, $n_\gamma$, within the GUP-modified thermodynamic framework and the constrained the GUP parameter $\beta_0$. This allows us to identify the baryon asymmetry $\eta$ generated in the early universe.

Our findings suggest that GUP-induced modifications to the thermodynamic properties of the primordial cosmic fluid can provide a viable framework for gravitational baryogenesis. In particular, the departure of the equation-of-state parameter from its conventional radiation-dominated value, arising from the modified phase-space measure, offers a viable mechanism for generating a nonvanishing baryon asymmetry in the early universe. This establishes an interesting connection between quantum gravity motivated modifications to statistical thermodynamics and the origin of the matter-antimatter asymmetry.

\section*{Acknowledgements}
Surendra Kumar Gour is supported through a Research Fellowship from the Ministry of Education, Government of India.

\paragraph{Funding:}
The authors received no specific funding for this work.

\paragraph{Competing Interests:}
The authors declare that they have no competing and/or financial interests.

\paragraph{Data Availability:}
No datasets were generated or analysed in this study.

\appendix

\section*{APPENDIXES}

\section{Baryonic Number Density Integral \boldmath $I_1(a)$}
\label{sec-A}

In this appendix, we evaluate the baryonic number density integral (\ref{eq-I1}), expressed as
\begin{equation}
 I_1(a) = \int_0^\infty \frac{e^x}{(e^x+1)^2} \frac{x^2 }{\left(a^2+x^2\right)^3} dx.
 \label{eq-I11}
\end{equation}

Since the integrand  in (\ref{eq-I11}) is a product of two functions, having the general form
\begin{equation}
 I(a) = \int_0^\infty dx\,  f(x)\,  g(x),
 \label{eq-f1g1}
\end{equation}
it is convenient to evaluate such integrals via performing convolution of their Mellin transforms.

Writing the Mellin transforms \cite{Sneddon-1972, Erdelyi-1954, Paris-2001} of the functions $f(x)$ and $g(x)$ as
\begin{equation}
   F(s) =  \int_0^\infty x^{s-1}f(x)\,dx \quad \text{and} \quad G(s)=  \int_0^\infty x^{s-1}g(x)\,dx,
\end{equation}
Mellin’s Parseval (convolution) theorem says
\begin{equation}
    I(a) = \int_{0}^{\infty} \,f(x)\,g(x)\,dx=\frac{1}{2\pi i} \int_{c-i\infty}^{c+i\infty}F(s)\,G(1-s)\,ds,
\label{eq-convo}    
\end{equation}
where the contour integration is to be carried out by evaluating the residues from the poles on the complex $s$ plane.

Defining
\begin{equation}
    f(x)=\frac{e^x}{(e^x+1)^2} \quad \text{and} \quad g(x)= \frac{x^2 }{\left(a^2+x^2\right)^3},
\end{equation}
their Mellin transforms are expressed as 
\begin{equation}
F(s)=\int_0^\infty x^{s-1} \frac{e^x}{(e^x+1)^2}\,dx \quad \text{and} \quad  G(s) = \int_{0}^{\infty} x^{s-1}\,\frac{x^2}{(a^2+x^2)^3}\,dx.
\label{eq-FG1}
\end{equation}

To evaluate $F(s)$ give by (\ref{eq-FG1}), we employ the expansion
\begin{equation}
\frac{e^x}{(e^x+1)^2}=\sum_{n=1}^{\infty}(-1)^{n-1} \,n e^{-nx},
\end{equation}
so that
\begin{equation}
F(s)= \sum_{n=1}^{\infty}(-1)^{n-1}n \int_0^\infty x^{s-1}e^{-nx}\,dx.
\end{equation}

Using the Gamma integral $\int_0^\infty x^{s-1}e^{-nx}\,dx=\frac{\Gamma(s)}{n^s}$, we get
\begin{equation}
F(s)=\Gamma(s)\sum_{n=1}^{\infty}\frac{(-1)^{n-1}}{n^{s-1}}= \Gamma(s)\left(1-2^{2-s}\right)\zeta(s-1),
\label{eq-F}
\end{equation}
since the sum reduces to the Dirichlet eta function $\eta(z)=\sum_{n=1}^{\infty}\frac{(-1)^{n-1}}{n^{z}}$, which is related to the Riemann zeta function via $\eta(z)=\left(1-2^{1-z}\right)\zeta(z)$.

To evaluate $G(s)$ given by (\ref{eq-FG1}), we substitute $x=a\sqrt y$, leading to
\begin{equation}
    G(s) = \frac{1}{2}a^{s-4} \int_{0}^{\infty} \frac{y^{s/2}}{(y+1)^3}\,dy.
\end{equation}

This integral is in the form of the beta function \cite{Arfken-1995}
\begin{equation}
    B(p,q) = \int_{0}^{\infty} \frac{t^{p-1}}{(1+t)^{p+q}}\, dt = \frac{\Gamma(p)\,\Gamma(q)}{\Gamma(p+q)},
\label{eq-beta}    
\end{equation}
leading to
\begin{equation}
    G(s)= \frac{a^s}{4a^4} \Gamma\left(\frac{2+s}{2}\right)\,\Gamma\left(\frac{4-s}{2}\right).
\label{eq-G}    
\end{equation}

Using Mellin’s Parseval theorem (\ref{eq-convo}), and substituting from equations (\ref{eq-F}) and (\ref{eq-G}), we find the integral $I_1(a)$ (\ref{eq-I11}) through the convolution 
\begin{equation}
I_1(a)  = \frac{1}{2\pi i} \int_{c-i\infty}^{c+i\infty} \Gamma(s)\zeta(s-1) \Gamma\left(\frac{3-s}{2}\right) \Gamma\left(\frac{3+s}{2}\right) \left(1-2^{2-s}\right)\frac{a^{-(s+3)}}{4}\,ds.
\end{equation}

A suitable choice for the contour can be made in the region $0 < {\rm Re}\ s < 3$ where this complex integral is well-defined. 

We evaluate this integral by employing Cauchy's residue theorem. Noting that there exists a series of poles at $s=0,-1,-2,-3,-4, \cdots$, we calculate the residues at these poles.

{\bf Residue at \boldmath $s=0$.} Noting that $\Gamma(s)$ has a simple pole at $s=0$, we write $s=0+ \varepsilon$, so that we have the expansion
\begin{equation}
\Gamma(0+\varepsilon) = \frac{1}{\varepsilon} - \gamma + \left( \frac{\gamma^2}{2} + \frac{\pi^2}{12} \right) \varepsilon+ \mathcal{O}(\varepsilon^2),
\end{equation}
where $\gamma$ is the Euler-Mascheroni constant $(\gamma \approx 0.57721)$. Consequently, the residue at $s=0$ is obtained as 
\begin{align}
 \operatorname*{Res}_{s=0}I_1(a)&=\lim_{\varepsilon \to 0}  \left[ \varepsilon \left(\frac{1}{\varepsilon} - \gamma \right) \zeta(-1) \Gamma\left(\frac{3}{2}\right) \Gamma\left(\frac{3}{2}\right) \left(1-2^{2}\right)\frac{a^{-3}}{4} \right] \nonumber \\&
 =(1)\left(-\frac1{12}\right)\left(\frac{\sqrt{\pi}}2\right)^2  (-3) \frac{a^{-3}}{4}=\frac{\pi}{64 a^3}.
\label{eq-R10} 
\end{align}

{\bf Residue at \boldmath $s=-1$.} For the pole at $s=-1$, writing $s=-1 +\varepsilon $ reveals the pole of $ \Gamma(s)$ via the expansion $\Gamma(-1+\varepsilon)
=-\frac1{\varepsilon}+\mathcal{O}(1)$. However, since $\zeta(-2)=0$ multiples the integrand, there is no contribution from this pole. 

{\bf Residue at \boldmath $s=-2$.} For the pole at $s=-2$, writing $s=-2 +\varepsilon $ reveals the pole of $ \Gamma(s)$ via the expansion $\Gamma(-2+\varepsilon)  = \frac{1}{2\varepsilon} + \left(\frac{3}{4} - \frac{\gamma}{2}\right) +  \mathcal{O}(\varepsilon)$. Consequently, the residue at $s=-2$ is given by 
\begin{align}
 \operatorname*{Res}_{s=-2}I_1(a)&=\lim_{\varepsilon \to 0} \left[  \varepsilon \left(\frac{1}{2\varepsilon} +  \mathcal{O}(1) \right) \zeta(-3) \Gamma\left(\frac{5}{2}\right) \Gamma\left(\frac{1}{2}\right) \left(1-2^{4}\right)\frac{a^{-1}}{4} \right] \nonumber \\&
 =\frac{1}{2}\frac{1}{120}\frac{3\sqrt{\pi}}{4}\sqrt{\pi}(-15) \frac{a^{-1}}{4}=-\frac{3\pi}{256a}.
 \label{eq-R12} 
\end{align}

{\bf Residue at \boldmath $s=-3$.} At $s=-3$, there is a double-pole coming from $\Gamma(s)$ and $\Gamma\left(\frac{3+s}{2}\right)$. Writing $s=-3 +\varepsilon $, the double-pole is revealed by the expansions $\Gamma(-3+\varepsilon)
=-\frac1{6 \varepsilon}+\mathcal{O}(1)$ and $\Gamma\left(\frac{\varepsilon}{2}\right) =\frac{2}{\varepsilon}-\gamma+\mathcal{O}(\varepsilon)$. However, although $\zeta(-4)=0$ multiples the integrand, there is now a non-zero contribution from this double-pole, given by
\begin{equation}
\mathop{\rm Res}_{s=-3} = \lim_{s\to-3}\,
\frac{d}{ds}\left[ (s+3)^2\,
\Gamma(s)\zeta(s-1) \Gamma\left(\frac{3-s}{2}\right) \Gamma\left(\frac{3+s}{2}\right) \left(1-2^{2-s}\right)\frac{a^{-(s+3)}}{4} \right],
\end{equation}
so that
\begin{equation}
\mathop{\rm Res}_{s=-3} = \lim_{\varepsilon\to0}\,
\frac{d}{d\varepsilon}\left[ \varepsilon^2\,
\Gamma(-3+\varepsilon)\zeta(-4+\varepsilon)
\Gamma\left(\frac{6-\varepsilon}{2}\right)
\Gamma\left(\frac{\varepsilon}{2}\right) \left(1-2^{5-\varepsilon}\right)
\frac{a^{-\varepsilon}}{4} \right].
\end{equation}

Since the leading order terms in the expansions contribute to this limit, we have 
\begin{equation}
\mathop{\rm Res}_{s=-3} = \lim_{\varepsilon\to0}\,
\frac{d}{d\varepsilon}\left[ \varepsilon^2\, \left(-\frac{1}{6\varepsilon}\right) \left\{\varepsilon\zeta'(-4) \right\} (2) \left(\frac{2}{\varepsilon}\right) (-31)
\frac{1}{4} \right] = \frac{31\zeta(5)}{8\pi^4},
\label{eq-R13} 
\end{equation}
 since $\zeta'(-4)=\frac{3}{4\pi^4}\zeta(5)$.
 
{\bf Residue at \boldmath $s=-4$.} For the pole at $s=-4$, writing $s=-4 +\varepsilon $ reveals the pole of $ \Gamma(s)$ via the expansion $\Gamma(-4+\varepsilon)  = \frac{1}{24\varepsilon} + \left(\frac{25}{288} - \frac{\gamma}{24}\right) +  \mathcal{O}(\varepsilon)$. Consequently, the residue at $s=-4$ is given by 
\begin{align}
 \operatorname*{Res}_{s=-4}I_1(a)&=\lim_{\varepsilon \to 0} \left[\varepsilon \left(\frac{1}{24\varepsilon} +  \mathcal{O}(1) \right) \zeta(-5) \Gamma\left(\frac{7}{2}\right) \Gamma\left(-\frac{1}{2}\right) \left(1-2^{6}\right)\frac{a}{4}\right] \nonumber \\&
 =\frac{1}{24}\left(-\frac{1}{252}\right)\frac{15\sqrt{\pi}}{8}(-2\sqrt{\pi})(-63)\frac{a}{4}=-\frac{5\pi a}{512}.
 \label{eq-R14} 
\end{align}

{\bf Total Contribution to \boldmath $I_1(a)$.} Adding the residues obtained in equations (\ref{eq-R10}), (\ref{eq-R12}), (\ref{eq-R13}) and (\ref{eq-R14}), we obtain 
\begin{equation}
    I_1(a)=\frac{\pi}{64a^3}-\frac{3\pi}{256a}+\frac{31\zeta(5)}{8\pi^4}-\frac{5\pi a}{512}+\cdots.
\end{equation}

Since $a \ll 1$, it is only the first term in the expansion that contributes dominantly.

\section{Photon Number Density Integral \boldmath $I_2(a)$}
\label{sec-B}

In this appendix, we evaluate the photon number density integral (\ref{eq-I2}), expressed as
\begin{equation}
I_2(a) = \int_0^\infty \frac{1}{(e^x-1)} \frac{x^2 }{\left(a^2+x^2\right)^3} dx.
\label{eq-I22}
\end{equation}

Defining
\begin{equation}
    f(x)=\frac{1}{(e^x-1)} \quad \text{and} \quad g(x)= \frac{x^2 }{\left(a^2+x^2\right)^3},
\end{equation}
their Mellin transforms are expressed as 
\begin{equation}
F(s)=\int_0^\infty x^{s-1} \frac{1}{(e^x-1)}\,dx \quad \text{and} \quad  G(s) = \int_{0}^{\infty} x^{s-1}\,\frac{x^2}{(a^2+x^2)^3}\,dx.
\label{eq-FG2}
\end{equation}

The Mellin transform  $F(s)$ immediately turns out to be
\begin{equation}
F(s) = \int_{0}^{\infty}x^{s-1}\,\frac{1}{e^x-1}\,dx= \Gamma(s)\zeta(s),
\label{eq-F2}
\end{equation}
whereas the Mellin transform $G(s)$ has already been evaluated in equation (\ref{eq-G}).

Using Mellin’s Parseval theorem  (\ref{eq-convo}), and substituting from equations (\ref{eq-F2}) and (\ref{eq-G}), we find the integral $I_2(a)$ given in (\ref{eq-I22}) through the convolution 
\begin{equation}
I_2(a)  = \frac{1}{2\pi i} \int_{c-i\infty}^{c+i\infty} \Gamma(s)\zeta(s)\Gamma\left(\frac{3-s}{2}\right)
\Gamma\left(\frac{3+s}{2}\right)\frac{a^{-(s+3)}}{4}\,ds.
\end{equation}

We choose the contour in the regular region $1 < {\rm Re}\ s <3$ and evaluate this integral employing Cauchy's residue theorem. Consequently, we obtain residues at the poles $s=1,0,-1,-2,-3,-4, \cdots$.

{\bf Residue at \boldmath $s=1$.}  The first residue comes from the simple pole of $\zeta(s)$ at $s=1$. Writing $s=1+ \varepsilon$, we have
\begin{align}
\mathop{\rm Res}_{s=1} I_2(a) &= \lim_{\varepsilon\to0}\,\left[ \varepsilon\,\Gamma(1+\varepsilon)\zeta(1+\varepsilon)
\Gamma\left(\frac{2-\varepsilon}{2}\right)\Gamma\left(\frac{4+\varepsilon}{2}\right)\frac{a^{-4-\varepsilon}}{4} \right],
\end{align}
upon using the expansion of $\zeta(s)$ about $s=1$,
\begin{equation}
\zeta(1+\varepsilon) = \frac{1}{\varepsilon} + \gamma - \gamma_1 \varepsilon + \frac{\gamma_2}{2!} \varepsilon^2 + \mathcal{O}(\varepsilon^3)
\end{equation}
where $\gamma_n$ are higher order Euler-Mascheroni constants.
\begin{align}
\mathop{\rm Res}_{s=1} I_2(a) &= \lim_{\varepsilon\to0}\,\left[ \varepsilon\,\Gamma(1)\left(\frac{1}{\varepsilon} + \mathcal{O}(1) \right) \Gamma\left(\frac{2}{2}\right)\Gamma\left(\frac{4}{2}\right)\frac{a^{-4}}{4} \right] \nonumber \\& = \Gamma(1)\Gamma(1)
\Gamma(2)\frac{a^{-4}}{4} =\frac{1}{4a^4}.
\label{eq-R211}
\end{align}

{\bf Residue at \boldmath $s=0$.}  The second residue comes from the pole of $\Gamma(s)$ at $s=0$, by writing $s=0+ \varepsilon$, we get
\begin{equation}
\mathop{\rm Res}_{s=0} I_2(a)= \lim_{\varepsilon\to0}\,\left[  \varepsilon\,\Gamma(\varepsilon)\zeta(\varepsilon)\Gamma\left(\frac{3-\varepsilon}{2}\right)\Gamma\left(\frac{3+\varepsilon}{2}\right)
\frac{a^{-3-\varepsilon}}{4} \right].
\end{equation}
Expanding $\Gamma(s)$ about $s=0$, we have
\begin{equation}
\Gamma(\varepsilon) = \frac{1}{\varepsilon} - \gamma + \left( \frac{\gamma^2}{2} + \frac{\pi^2}{12} \right) \varepsilon+ \mathcal{O}(\varepsilon^2),
\end{equation}
so that
\begin{align}
\mathop{\rm Res}_{s=0} I_2(a)&= \lim_{\varepsilon\to0}\,\left[  \varepsilon\,\left(\frac{1}{\varepsilon} + \mathcal{O}(1)\right)\zeta(0)\Gamma\left(\frac{3}{2}\right)\Gamma\left(\frac{3}{2}\right)\frac{a^{-3-\varepsilon}}{4} \right] \nonumber \\& = -\frac12 \frac{\sqrt{\pi}}{2} \frac{\sqrt{\pi}}{2} \frac{a^{-3}}{4} = -\frac{\pi}{32a^3}.
\label{eq-R20}
\end{align}

{\bf Residue at \boldmath $s=-1$.} The third residue comes from the pole of $\Gamma(s)$ at $s=-1$. Writing $s=-1+ \varepsilon$,
\begin{equation}
\mathop{\rm Res}_{s=-1} I_2(a) = \lim_{\varepsilon\to0}\,
\left[ \varepsilon\,\Gamma(-1+\varepsilon)\zeta(-1+\varepsilon)\Gamma\left(\frac{4-\varepsilon}{2}\right)\Gamma\left(\frac{2+\varepsilon}{2}\right) \frac{a^{-2-\varepsilon}}{4} \right],
\end{equation}
upon using the expansion of $\Gamma(s)$ about $s=-1$, given by
\begin{equation}
\Gamma(-1+\varepsilon) = -\frac{1}{\varepsilon} + (\gamma - 1) - \left( \frac{\gamma^2}{2} - \gamma + 1 + \frac{\pi^2}{12} \right) \varepsilon + \mathcal{O}(\varepsilon^2).
\end{equation}
Consequently, 
\begin{align}
\mathop{\rm Res}_{s=-1} I_2(a) &= \lim_{\varepsilon\to0}\,
\left[ \varepsilon\, \left(-\frac{1}{\varepsilon} + \mathcal{O}(1)\right)  \zeta(-1)\Gamma\left(\frac{4}{2}\right)\Gamma\left(\frac{2}{2}\right) \frac{a^{-2}}{4} \right] \nonumber \\&  = -\left(-\frac{1}{12} \frac{a^{-2}}{4}\right) = \frac{1}{48a^2}.
\label{eq-R21}
\end{align}

{\bf Residue at \boldmath $s=-2$.} The fourth residue comes from the pole of $\Gamma(s)$ at $s=-2$. Writing $s=-2+ \varepsilon$,
\begin{equation}
\mathop{\rm Res}_{s=-2} I_2(a) = \lim_{\varepsilon\to0}\,
\left[ \varepsilon\,\Gamma(-2+\varepsilon)\zeta(-2+\varepsilon)\Gamma\left(\frac{5-\varepsilon}{2}\right)\Gamma\left(\frac{1+\varepsilon}{2}\right)\frac{a^{-1-\varepsilon}}{4} \right] 
\end{equation}
Noting the expansion of $\Gamma(s)$ about $s=-2$,
\begin{equation}
\Gamma(-2+\varepsilon) = \frac{1}{2\varepsilon} + \left(\frac{3}{4} - \frac{\gamma}{2}\right) + \left( \frac{\gamma^2}{4} - \frac{3\gamma}{4} + \frac{7}{8} + \frac{\pi^2}{24} \right) \varepsilon + \mathcal{O}(\varepsilon^2),
\end{equation}
we see that the residue vanishes, $\mathop{\rm Res}_{s=-2} I_2(a)=0$, because $\zeta(-2) = 0$.

{\bf Residue at \boldmath $s=-3$.} The fifth residue comes from the double pole due to $\Gamma(s)$ and  $\Gamma\left(\frac{3+s}{2}\right)$ at $s=-3$. Writing $s=-3+\varepsilon$,
\begin{equation}
\mathop{\rm Res}_{s=-3}  I_2(a) = \lim_{s\to-3}\,
\frac{d}{ds}\left[ (s+3)^2\,
\Gamma(s)\zeta(s)
\Gamma\left(\frac{3-s}{2}\right)
\Gamma\left(\frac{3+s}{2}\right)
\frac{a^{-(s+3)}}{4} \right],
\end{equation}
so that
\begin{equation}
\mathop{\rm Res}_{s=-3}  I_2(a) = \lim_{\varepsilon\to0}\,
\frac{d}{d\varepsilon}\left[ \varepsilon^2\,
\Gamma(-3+\varepsilon)\zeta(-3+\varepsilon)
\Gamma\left(\frac{6-\varepsilon}{2}\right)
\Gamma\left(\frac{\varepsilon}{2}\right)
\frac{a^{-\varepsilon}}{4} \right].
\end{equation}
To evaluate this residue, we use the expansions 
\begin{equation}
\Gamma(-3+\varepsilon) = -\frac{1}{6\varepsilon} + A  + B\varepsilon + \mathcal{O}(\varepsilon^2), \quad
\zeta(-3+\varepsilon)  = \zeta(-3)+\varepsilon\zeta'(-3)+\mathcal{O}(\varepsilon^2),
\end{equation}
\begin{equation}
\Gamma\left(3 - \frac{\varepsilon}{2}\right) = 2 - \left(\frac{3}{2} - \gamma\right) \varepsilon + \mathcal{O}(\varepsilon^2), \quad
\Gamma\left(\frac{\varepsilon}{2}\right) = \frac{2}{\varepsilon} - \gamma + C \varepsilon + \mathcal{O}(\varepsilon^2),
\end{equation}
and
\begin{equation}
a^{-\varepsilon}=e^{-\varepsilon\ln a}=1-\varepsilon\ln a + \varepsilon^2\ln^2 a + \mathcal{O}(\varepsilon^3),
\end{equation}
where
\begin{equation}
A=\left( \frac{\gamma}{6} - \frac{11}{36} \right), \quad
B=- \left( \frac{\gamma^2}{12} - \frac{11\gamma}{36} + \frac{85}{216} + \frac{\pi^2}{72} \right), \quad C=\frac{1}{4}\left(\gamma^2 + \frac{\pi^2}{3}\right),
\end{equation}
leading to
\begin{align}
\operatorname{Res}_{s=-3} I_2(a)
&= \lim_{\varepsilon\to0}\frac{d}{d\varepsilon}
\Bigg[\varepsilon^2\left(-\frac{1}{6\varepsilon}+A+B\varepsilon\right)
\left\{\zeta(-3)+\varepsilon\zeta'(-3)\right\} \\
&\quad
\left\{2-\left(\frac32-\gamma\right)\varepsilon\right\}
\left(\frac{2}{\varepsilon}-\gamma\right)
\left(\frac{1-\varepsilon\ln a}{4}\right)
\Bigg].
\end{align}
Since only the leading order terms in the expansions contribute to the above limit, we have
\begin{equation}
\operatorname{Res}_{s=-3} I_3(a)=-\frac{\zeta'(-3)}{6}-\frac{13}{8640}+\frac{\gamma}{720}+\frac{\ln a}{720}.
\label{eq-R23}
\end{equation}

{\bf Residue at \boldmath $s=-4$.} The sixth residue comes from the pole of $\Gamma(s)$ at $s=-4$. Thus, although the pole of $\Gamma(-4+\varepsilon)$ is cancelled, the residue vanishes, $\mathop{\rm Res}_{s=-4} I_2(a)=0$, because $\zeta(-4) = 0$.

{\bf Total Contribution to \boldmath $I_2(a)$.} Adding the residues obtained in equations (\ref{eq-R211}), (\ref{eq-R20}), (\ref{eq-R21}) and (\ref{eq-R23}) 
\begin{equation}
I_2(a)= \frac{1}{4a^4}-\frac{\pi}{32a^3}+\frac{1}{48a^2}-\frac{\zeta'(-3)}{6}-\frac{13}{8640}+\frac{\gamma}{720}+\frac{\ln a}{720} + \cdots.
\end{equation}

\section{Bosonic Energy Density Integral \boldmath $I_3(a)$}
\label{sec-C}

In this appendix, we evaluate the bosonic radiation energy density integral (\ref{eq-I3}), expressed as
\begin{equation}
    I_3(a) = \int_0^\infty \frac{x^3dx}{(e^x-1)(a^2+x^2)^3}.
\label{eq-I33}    
\end{equation}

We define an auxiliary function
\begin{equation}
    J(a) = \int_0^\infty \frac{x^3dx}{(e^x-1)(a^2+x^2)}.
\label{eq-aux}    
\end{equation}
Differentiating twice with respect to $a$, we obtain
\begin{equation}
    J'(a) = -2a \int_0^\infty \frac{x^3dx}{(e^x-1)(a^2+x^2)^2},
\label{eq-auxd}    
\end{equation}
\begin{equation}
    J''(a) = -2 \int_0^\infty \frac{x^3dx}{(e^x-1)(a^2+x^2)^2} + 8a^2 \int_0^\infty \frac{x^3dx}{(e^x-1)(a^2+x^2)^3}.
\label{eq-auxdd}    
\end{equation}
 
Combining (\ref{eq-I33}), (\ref{eq-auxd}) and (\ref{eq-auxdd}), we get a relation between the integral and the derivatives of the auxiliary function as
\begin{equation}
I_3(a) = \frac{1}{8a^2} \, J''(a) - \frac{1}{8a^3} \, J'(a).
\label{eq-I3T}
\end{equation}

Since $ \frac{x^3}{a^2+x^2} = x - \frac{a^2x}{a^2+x^2}$, we can write the auxiliary integral (\ref{eq-aux})
in the form 
\begin{equation}
    J(a) = \int_0^\infty \frac{xdx}{(e^x-1)} - a^2 \int_0^\infty \frac{xdx}{(e^x-1)(a^2+x^2)}.
\end{equation}

The first integral can be immediately evaluated as
\begin{equation}
    \int_0^\infty \frac{xdx}{(e^x-1)} = \Gamma(2) \zeta(2) = \frac{\pi^2}{6}.
\end{equation}
For the second integral, we employ Binet's second representation of the digamma function \cite{Erdelyi-1953, Whittaker-1996, NIST-2010},
\begin{equation}
    \psi(z) = \ln z -\frac{1}{2z} -2 \int_0^\infty \frac{tdt}{(t^2+z^2)(e^{2\pi t}-1)}.
\end{equation}
Setting $2\pi t=x$ and $2\pi z=a$, we have
\begin{equation}
    \psi\left(\textstyle\frac{a}{2\pi}\right) = \ln \left(\textstyle\frac{a}{2\pi}\right) -\frac{\pi}{a} -2 \int_0^\infty \frac{xdx}{(a^2+x^2)(e^{x}-1)},
\end{equation}
so that
\begin{equation}
    \int_0^\infty \frac{xdx}{(a^2+x^2)(e^{x}-1)} = \frac{1}{2} \left[ \ln \left(\textstyle\frac{a}{2\pi}\right)- \frac{\pi}{a} - \psi\left(\textstyle\frac{a}{2\pi}\right) \right].
\end{equation}

Substituting this expression in the auxiliary integral (\ref{eq-aux}), we get
\begin{equation}
    J(a) = \frac{\pi^2}{6} - \frac{a^2}{2} \left[ \ln \left(\textstyle\frac{a}{2\pi}\right) - \frac{\pi}{a}- \psi\left(\textstyle\frac{a}{2\pi}\right)\right].
\label{eq-Jf}    
\end{equation}

Differentiating \ref{eq-Jf} with respect to $a$, we have 
\begin{equation}
    J'(a) = \frac{\pi}{2} - \frac{a}{2}
- a \ln \left(\textstyle\frac{a}{2\pi}\right)
+ a \psi\left(\textstyle\frac{a}{2\pi}\right)
+ \left(\textstyle\frac{a^2}{4\pi}\right) \psi'\left(\textstyle\frac{a}{2\pi}\right),
\end{equation}
\begin{equation}
    J''(a) = -\frac{3}{2} 
- \ln \left(\textstyle\frac{a}{2\pi}\right)
+ \psi\left(\textstyle\frac{a}{2\pi}\right)
+ \left(\textstyle\frac{a}{\pi}\right) \psi'\left(\textstyle\frac{a}{2\pi}\right)
+ \frac{1}{2}\left(\textstyle\frac{a}{2\pi}\right)^2 \psi''\left(\textstyle\frac{a}{2\pi}\right).
\end{equation}

Substituting the above results in (\ref{eq-I3T}), we obtain
\begin{equation}
    I_3(a) =  - \frac{\pi}{16a^3}
-\frac{1}{8a^2}
+ \left(\textstyle\frac{3}{32\pi a}\right) \psi'\left(\textstyle\frac{a}{2\pi}\right)
+ \left(\textstyle\frac{1}{64\pi^2}\right) \psi''\left(\textstyle\frac{a}{2\pi}\right).
\label{eq-I3TF}
\end{equation}

Employing the expansions \cite{Erdelyi-1953, Abramowitz-1964, Olver-2010}
\begin{equation}
\psi(z) = -\frac{1}{z} - \gamma+ \zeta(2)z - \zeta(3)z^2+ \zeta(4)z^3 + \cdots,
\end{equation}
\begin{equation}
\psi'(z) = \frac{1}{z^2} + \zeta(2) - 2 \zeta(3)z+ 3\zeta(4)z^2 + \cdots,
\end{equation}
\begin{equation}
\psi''(z) = -\frac{2}{z^3}  - 2\zeta(3)+ 6\zeta(4)z + \cdots,
\end{equation}
in (\ref{eq-I3TF}), we finally obtain
\begin{equation}
I_3(a)= \frac{\pi}{16a^3}-\frac{1}{8a^2}+\frac{3\zeta(2)}{32 \pi a} - \frac{\zeta(3)}{8\pi^2} + \frac{21 \zeta(4) a}{64 \pi^3} + \cdots.
\end{equation}

\section{Bosonic Pressure Integral \boldmath $I_4(a)$}
\label{sec-E}

In this appendix, we evaluate the bosonic radiation pressure integral (\ref{eq-I6}), expressed by
\begin{equation}
I_4(a)=\int_0^\infty \ln\left(1-e^{-x}\right) \frac{x^2}{(a^2+x^2)^3} dx.
\label{eq-I55}
\end{equation}

Defining
\begin{equation}
    f(x)=\ln\left(1-e^{-x}\right) \quad \text{and} \quad g(x)= \frac{x^2 }{\left(a^2+x^2\right)^3},
\end{equation}
their Mellin transforms are expressed as 
\begin{equation}
F(s)=\int_0^\infty x^{s-1} \ln\left(1-e^{-x}\right)\,dx \quad \text{and} \quad  G(s) = \int_{0}^{\infty} x^{s-1}\,\frac{x^2}{(a^2+x^2)^3}\,dx.
\label{eq-FG5}
\end{equation}

The Mellin transform  $F(s)$ turns out to be
\begin{equation}
F(s) = \int_{0}^{\infty}x^{s-1}\,\ln\left(1-e^{-x}\right)\,dx= -\Gamma(s) \zeta(s+1),
\label{eq-F5}
\end{equation}
using the expansion $\ln\left(1-e^{-x}\right)= -\sum_{n=1}^{\infty}\frac{e^{-nx}}{n}$ and the relation between Riemann zeta-function and the Dirichlet eta-function.  The Mellin transform $G(s)$ is given equation (\ref{eq-G}).

Employing Mellin’s Parseval theorem (\ref{eq-convo}), and substituting from equations (\ref{eq-F5}) and (\ref{eq-G}), we find the integral $I_4(a)$ given in (\ref{eq-I55}) through 
\begin{equation}
I_4(a)  = - \frac{1}{2\pi i} \int_{c-i\infty}^{c+i\infty} \Gamma(s)\zeta(s+1)\, \Gamma\left(\frac{3-s}{2}\right)\Gamma\left(\frac{3+s}{2}\right)\frac{a^{-(s+3)}}{4}\,ds.
\end{equation}

Choosing the contour in the regular region $0 < {\rm Re}\ s <3$, we evaluate this integral by picking up the poles $s=0,-1,-2,-3,-4, \cdots$.

{\bf Residue at \boldmath $s=0$.} At $s=0$, there is a double-pole coming from $\Gamma(s)$ and $\zeta(s+1)$. Writing $s=0 +\varepsilon $, the double-pole occurs by the expansions 
\begin{equation}
\Gamma(0+\varepsilon) = \frac{1}{\varepsilon} - \gamma + \left( \frac{\gamma^2}{2} + \frac{\pi^2}{12} \right) \varepsilon+ \mathcal{O}(\varepsilon^2), \quad \zeta(1+\varepsilon) = \frac{1}{\varepsilon} + \gamma - \gamma_1 \varepsilon + \frac{\gamma_2}{2!} \varepsilon^2 + \mathcal{O}(\varepsilon^3),
\end{equation}
and
\begin{equation}
 a^{-(3+\varepsilon)} =a^{-3}-\varepsilon a^{-3} \ln a+\frac{\varepsilon^2(\ln a)^2}{2!} a^{-3}- \mathcal{O}(\varepsilon^3).
\end{equation}

The residue at $s=0$ is given by 
\begin{equation}
\mathop{\rm Res}_{s=0}  I_4(a) = -\lim_{s\to0}\,
\frac{d}{ds}\left[ (s+0)^2\,
\Gamma(s)\zeta(s+1) \Gamma\left(\frac{3-s}{2}\right)
\Gamma\left(\frac{3+s}{2}\right)\frac{a^{-(s+3)}}{4} \right],
\end{equation}
leading to 
\begin{equation}
\mathop{\rm Res}_{s=0}  I_4(a) = -\lim_{\varepsilon\to0}\,
\frac{d}{d\varepsilon}\left[ \varepsilon^2\,
\Gamma(\varepsilon)\zeta(1+\varepsilon) \Gamma\left(\frac{3-\varepsilon}{2}\right)
\Gamma\left(\frac{3+\varepsilon}{2}\right)\frac{a^{-(\varepsilon+3)}}{4}  \right].
\end{equation}
Consequently, we get
\begin{align}
\mathop{\rm Res}_{s=0}  I_4(a) =& -\lim_{\varepsilon\to0}\,
\frac{d}{d\varepsilon}\left[ \varepsilon^2\,
\left(\frac{1}{\varepsilon} - \gamma \right) \left(\frac{1}{\varepsilon} + \gamma\right) \Gamma\left(\frac{3-\varepsilon}{2}\right)
\Gamma\left(\frac{3+\varepsilon}{2}\right)\frac{(a^{-3}-\varepsilon a^{-3} \ln a)}{4}  \right] \nonumber \\& =  -\left(\frac{\sqrt{\pi}}{2}\right) \left(\frac{\sqrt{\pi}}{2}\right) \left(- \frac{\varepsilon a^{-3} \ln a}{4}\right)= \frac{\pi \ln a}{16a^3}.
\label{eq-R50}
\end{align}

{\bf Residue at \boldmath $s=-1$.} For the pole at $s=-1$, writing $s=-1 +\varepsilon $ introduces the pole of $ \Gamma(s)$ via the expansion $\Gamma(-1+\varepsilon)
=-\frac1{\varepsilon}+\mathcal{O}(1)$. Consequently, the residue at $s=-1$ is obtained as
\begin{align}
 \operatorname*{Res}_{s=-1}I_4(a)&= - \lim_{\varepsilon \to 0}  \left[ \varepsilon \left(-\frac{1}{\varepsilon} \right) \zeta(\varepsilon) \Gamma\left(\frac{4-\varepsilon}{2}\right) \Gamma\left(\frac{2+\varepsilon}{2}\right) \frac{a^{-(2+\varepsilon)}}{4} \right] \nonumber \\&
 =-\left(-1 \right)   \left(-\frac{1}{2}\right) \left(1 \right) \left( 1 \right) \frac{a^{-2}}{4}=-\frac{1}{8 a^2}.
\label{eq-R51} 
\end{align}

{\bf Residue at \boldmath $s=-2$.} For the pole at $s=-2$, writing $s=-2 +\varepsilon $ reveals the pole of $ \Gamma(s)$ via the expansion $\Gamma(-2+\varepsilon)
=\frac1{2\varepsilon}+\mathcal{O}(1)$. Consequently, the residue at $s=-2$ is given by 
\begin{align}
 \operatorname*{Res}_{s=-2}I_4(a)&=  -\lim_{\varepsilon \to 0}  \left[ \varepsilon \left(\frac{1}{2\varepsilon} \right) \zeta(-1+\varepsilon) \Gamma\left(\frac{5-\varepsilon}{2}\right) \Gamma\left(\frac{1+\varepsilon}{2}\right) \frac{a^{-(1+\varepsilon)}}{4} \right] \nonumber \\&
 =-\left(\frac{1}{2} \right)   \left(-\frac{1}{12}\right) \left(\frac{3\sqrt{\pi}}{4} \right) \left( \sqrt{\pi} \right) \frac{a^{-1}}{4}=\frac{ \pi}{128 a}.
\label{eq-R52} 
\end{align}

{\bf Residue at \boldmath $s=-3$.} At $s=-3$, there is a double-pole coming from $\Gamma(s)$ and $\Gamma\left(\frac{3+s}{2}\right)$. Writing $s=-3+\varepsilon $, the double-pole can be accounted for by the expansions 
\begin{equation}
\Gamma(-3+\varepsilon) = -\frac{1}{6\varepsilon} + A  + B\varepsilon + \mathcal{O}(\varepsilon^2), \quad
\zeta(-2+\varepsilon)  = \varepsilon\zeta'(-2)+\mathcal{O}(\varepsilon^2),
\end{equation}
\begin{equation}
\Gamma\left(3 - \frac{\varepsilon}{2}\right) = 2 - \left(\frac{3}{2} - \gamma\right) \varepsilon + \mathcal{O}(\varepsilon^2), \quad
\Gamma\left(\frac{\varepsilon}{2}\right) = \frac{2}{\varepsilon} - \gamma + C \varepsilon + \mathcal{O}(\varepsilon^2),
\end{equation}
and
\begin{equation}
a^{-\varepsilon}=e^{-\varepsilon\ln a}=1-\varepsilon\ln a + \varepsilon^2\ln^2 a + \mathcal{O}(\varepsilon^3),
\end{equation}
where
\begin{equation}
A=\left( \frac{\gamma}{6} - \frac{11}{36} \right), \quad
B=- \left( \frac{\gamma^2}{12} - \frac{11\gamma}{36} + \frac{85}{216} + \frac{\pi^2}{72} \right), \quad C=\frac{1}{4}\left(\gamma^2 + \frac{\pi^2}{3}\right).
\end{equation}

Importantly, although $\zeta{(-2)}=0$, the contribution from this double-pole is non-zero. The residue at $s=-3$ is found by 
\begin{equation}
\mathop{\rm Res}_{s=-3}  I_4(a) = -\lim_{s\to0}\,
\frac{d}{ds}\left[ (s+3)^2\,
\Gamma(s)\zeta(s+1) \Gamma\left(\frac{3-s}{2}\right)
\Gamma\left(\frac{3+s}{2}\right)\frac{a^{-(s+3)}}{4} \right],
\end{equation}
so that
\begin{equation}
\mathop{\rm Res}_{s=0}  I_4(a) = -\lim_{\varepsilon\to0}\,
\frac{d}{d\varepsilon}\left[ \varepsilon^2\,
\Gamma(-3+\varepsilon)\zeta(-2+\varepsilon) \Gamma\left(\frac{6-\varepsilon}{2}\right)
\Gamma\left(\frac{\varepsilon}{2}\right)
\frac{a^{-\varepsilon}}{4}  \right].
\end{equation}
Using $\zeta'(-2)=-\frac{\zeta(3)}{4 \pi^2}$, this leads to
\begin{align}
\mathop{\rm Res}_{s=-3}  I_4(a) =& -\lim_{\varepsilon\to0}\,
\frac{d}{d\varepsilon}\left[ \varepsilon^2\,
\left(-\frac{1}{6\varepsilon} \right) \left\{\varepsilon\zeta'(-2)\right\} \left(2\right)
\left(\frac{2}{\varepsilon}\right)\frac{(1)}{4}  \right] = -\frac{\zeta(3)}{24\pi^2}.
\label{eq-R53}
\end{align}

{\bf Residue at \boldmath $s=-4$.} For the pole at $s=-4$, we write $s=-4 +\varepsilon $, so that the pole of $ \Gamma(s)$ occurs via the expansion $\Gamma(-4+\varepsilon)
=\frac1{24\varepsilon}+\mathcal{O}(1)$, giving the residue 
\begin{align}
 \operatorname*{Res}_{s=-4}I_4(a)&=  -\lim_{\varepsilon \to 0}  \left[ \varepsilon \left(\frac{1}{24\varepsilon} \right) \zeta(-3+\varepsilon) \Gamma\left(\frac{7-\varepsilon}{2}\right) \Gamma\left(\frac{-1+\varepsilon}{2}\right) \frac{a^{(1-\varepsilon)}}{4} \right] \nonumber \\&
 =-\left(\frac{1}{24} \right)   \left(\frac{1}{120}\right) \left(\frac{15\sqrt{\pi}}{8} \right) \left(-2 \sqrt{\pi} \right) \frac{a}{4}=\frac{ \pi a}{3072}.
\label{eq-R54} 
\end{align}

{\bf Total Contribution to \boldmath $I_4(a)$.} Adding the residues obtained in equations (\ref{eq-R50}), (\ref{eq-R51}), (\ref{eq-R52}), (\ref{eq-R53})and (\ref{eq-R54}), we get
\begin{equation}
I_4(a)= \frac{\pi \ln a}{16a^3}-\frac{1}{8a^2}+\frac{\pi}{128 a} - \frac{\zeta(3)}{24\pi^2} + \frac{ \pi a}{3072} + \cdots.
\end{equation}

\section{Fermionic Energy Density Integral \boldmath $I_5(a)$}
\label{sec-D}

In this appendix, we evaluate the fermionic radiation energy density integral (\ref{eq-I4}), expressed by
\begin{equation}
    I_5(a) = \int_0^\infty  \frac{1}{e^x+1} \frac{x^3}{(a^2+x^2)^3} dx.
\label{eq-I44}    
\end{equation}

Defining
\begin{equation}
    f(x)=\frac{1}{(e^x+1)} \quad \text{and} \quad g(x)= \frac{x^3 }{\left(a^2+x^2\right)^3},
\end{equation}
the corresponding Mellin transforms are given by
\begin{equation}
F(s)=\int_0^\infty x^{s-1} \frac{1}{(e^x+1)}\,dx \quad \text{and} \quad  G(s) = \int_{0}^{\infty} x^{s-1}\,\frac{x^3}{(a^2+x^2)^3}\,dx.
\label{eq-FG4}
\end{equation}

To evaluate $F(s)$ in (\ref{eq-FG4}), we use $\frac{1}{(e^x+1)}= \sum_{n=1}^{\infty}(-1)^{n-1}e^{-nx}$, giving
\begin{equation}
F(s) = \sum_{n=1}^{\infty}(-1)^{n-1} \int_0^\infty x^{s-1}e^{-nx}\,dx.
\end{equation}

Using the Gamma integral $\int_0^\infty x^{s-1}e^{-nx}\,dx=\frac{\Gamma(s)}{n^s}$, and the zeta function, we get
\begin{equation}
F(s)= \Gamma(s)\sum_{n=1}^{\infty} \frac{(-1)^{n-1}}{n^s}  = \Gamma(s)(1-2^{1-s})\zeta(s).
\label{eq-F4}
\end{equation}

To evaluate $G(s)$ given by (\ref{eq-FG4}), we substitute $x=a \sqrt{y}$, leading to
\begin{equation}
    G(s) = \frac{1}{2}a^{s-3} \int_{0}^{\infty} \frac{y^{\frac{s+1}{2}}}{(y+1)^3}\,dy.
\end{equation}

Using the beta function (\ref{eq-beta}), we arrive at 
\begin{equation}
    G(s)= \frac{a^s}{4a^3} \Gamma\left(\frac{3+s}{2}\right)\,\Gamma\left(\frac{3-s}{2}\right).
\label{eq-G4}    
\end{equation}

Following Mellin’s Parseval theorem (\ref{eq-convo}), and substituting (\ref{eq-F4}) and (\ref{eq-G4}), the integral $I_4(a)$ given in (\ref{eq-I44}) is expressed as 
\begin{equation}
I_5(a)  = \frac{1}{2\pi i} \int_{c-i\infty}^{c+i\infty} \Gamma(s)\zeta(s) \Gamma\left(\frac{4-s}{2}\right) \Gamma\left(\frac{2+s}{2}\right) \left(1-2^{1-s}\right)\frac{a^{-(s+2)}}{4}\,ds.
\end{equation}

We choose the contour in the region $0 < {\rm Re}\ s < 4$ and evaluate this integral by calculating the residues at the poles $s=0,-1,-2,-3,-4, \cdots$.

{\bf Residue at \boldmath $s=0$.} Noting that $\Gamma(s)$ has a simple pole at $s=0$, we write $s=0+ \varepsilon$, and use the expansion
\begin{equation}
\Gamma(0+\varepsilon) = \frac{1}{\varepsilon} - \gamma + \left( \frac{\gamma^2}{2} + \frac{\pi^2}{12} \right) \varepsilon+ \mathcal{O}(\varepsilon^2).
\end{equation}
Consequently, the residue at $s=0$ is given by 
\begin{align}
 \operatorname*{Res}_{s=0}I_5(a)&=\lim_{\varepsilon \to 0}  \left[ \varepsilon \left(\frac{1}{\varepsilon} - \gamma \right) \zeta(0) \Gamma(2) \Gamma(1) \left(1-2\right)\frac{a^{-2}}{4} \right] \nonumber \\&
 =(1)\left(-\frac1{2}\right) (1) (1) (-1) \frac{a^{-2}}{4}=\frac{1}{8 a^2}.
\label{eq-R40} 
\end{align}

{\bf Residue at \boldmath $s=-1$.}  Writing $s=-1+ \varepsilon$, we have
\begin{equation}
\mathop{\rm Res}_{s=-1} I_5(a) = \lim_{\varepsilon\to0}\,
\left[ \varepsilon\,\Gamma(-1+\varepsilon)\zeta(-1+\varepsilon)\Gamma\left(\frac{5-\varepsilon}{2}\right)\Gamma\left(\frac{1+\varepsilon}{2}\right) \left(1-2^{2-\varepsilon}\right) \frac{a^{-1-\varepsilon}}{4} \right].
\end{equation}
Using the expansion 
\begin{equation}
\Gamma(-1+\varepsilon) = -\frac{1}{\varepsilon} + (\gamma - 1) - \left( \frac{\gamma^2}{2} - \gamma + 1 + \frac{\pi^2}{12} \right) \varepsilon + \mathcal{O}(\varepsilon^2),
\end{equation}
the residue at $s=-1$ is obtained as
\begin{align}
\mathop{\rm Res}_{s=-1} I_5(a) &= \lim_{\varepsilon\to0}\,
\left[ \varepsilon\, \left(-\frac{1}{\varepsilon} + \mathcal{O}(1)\right)  \zeta(-1)\Gamma\left(\frac{5}{2}\right)\Gamma\left(\frac{1}{2}\right)(-3) \frac{a^{-1}}{4} \right] \nonumber \\&  = (-1)
\left(-\frac1{12}\right)\left(\frac{3\sqrt{\pi}}4\right)(\sqrt{\pi})(-3) \frac{a^{-1}}{4}
 = -\frac{3 \pi}{64a}.
\label{eq-R41}
\end{align}

{\bf Residue at \boldmath $s=-2$.} At $s=-2$, there is a double-pole coming from $\Gamma(s)$ and $\Gamma\left(\frac{2+s}{2}\right)$. Writing $s=-2 +\varepsilon $, the double-pole is revealed by the expansions $\Gamma(-2+\varepsilon)
=\frac1{2 \varepsilon}+\mathcal{O}(1)$ and $\Gamma\left(\frac{\varepsilon}{2}\right) =\frac{2}{\varepsilon}-\gamma+\mathcal{O}(\varepsilon)$. However, although $\zeta(-2)=0$ multiples the integrand, there is a non-zero contribution from this double-pole, given by
\begin{equation}
\mathop{\rm Res}_{s=-2} I_5(a) = \lim_{s\to-2}\,
\frac{d}{ds}\left[ (s+2)^2\,
\Gamma(s)\zeta(s) \Gamma\left(\frac{4-s}{2}\right) \Gamma\left(\frac{2+s}{2}\right) \left(1-2^{1-s}\right)\frac{a^{-(s+2)}}{4} \right],
\end{equation}
so that
\begin{equation}
\mathop{\rm Res}_{s=-2} I_5(a) = \lim_{\varepsilon\to0}\,
\frac{d}{d\varepsilon}\left[ \varepsilon^2\,
\Gamma(-2+\varepsilon)\zeta(-2+\varepsilon)
\Gamma\left(\frac{6-\varepsilon}{2}\right)
\Gamma\left(\frac{\varepsilon}{2}\right) \left(1-2^{3-\varepsilon}\right)
\frac{a^{-\varepsilon}}{4} \right].
\end{equation}

Since the leading order terms in the expansions contribute to this limit, we have 
\begin{equation}
\mathop{\rm Res}_{s=-2} I_5(a)= \lim_{\varepsilon\to0}\,
\frac{d}{d\varepsilon}\left[ \varepsilon^2\, \left(\frac{1}{2\varepsilon}\right) \left\{\varepsilon\zeta'(-2) \right\} (2) \left(\frac{2}{\varepsilon}\right) (-7)
\frac{1}{4} \right] = \frac{7\zeta(3)}{8\pi^4},
\label{eq-R42}
\end{equation}
 since $\zeta'(-2)=-\frac{\zeta(3)}{4\pi^2}$.

{\bf Residue at \boldmath $s=-3$.}  Writing $s=-3+ \varepsilon$, we get
\begin{equation}
\mathop{\rm Res}_{s=-3} I_5a) = \lim_{\varepsilon\to0}\,
\left[ \varepsilon\,\Gamma(-3+\varepsilon)\zeta(-3+\varepsilon)\Gamma\left(\frac{7-\varepsilon}{2}\right)\Gamma\left(\frac{-1+\varepsilon}{2}\right) \left(1-2^{4-\varepsilon}\right) \frac{a^{1-\varepsilon}}{4} \right].
\end{equation}
Using the expansion $\Gamma(-3+\varepsilon) = -\frac{1}{6\varepsilon} + \mathcal{O}(1)$, the residue at $s=-3$ is obtained as
\begin{align}
\mathop{\rm Res}_{s=-3} I_5(a) &= \lim_{\varepsilon\to0}\,
\left[ \varepsilon\, \left(-\frac{1}{6\varepsilon} + \mathcal{O}(1)\right)  \zeta(-3)\Gamma\left(\frac{7}{2}\right)\Gamma\left(-\frac{1}{2}\right)(-15) \frac{a}{4} \right] \nonumber \\&  = \left(-\frac{1}{6}\right)
\left(\frac1{120}\right)\left(\frac{15\sqrt{\pi}}{8}\right)(-2\sqrt{\pi})(-15) \frac{a}{4}
 = -\frac{5 \pi a}{256}.
\label{eq-R43}
\end{align}

{\bf Residue at \boldmath $s=-4$.} At $s=-4$, although the pole of $\Gamma(-4+\varepsilon)$ is cancelled, the residue vanishes, $\mathop{\rm Res}_{s=-4} I_5(a)=0$, since $\zeta(-4) = 0$.

{\bf Total Contribution to \boldmath $I_5(a)$.} Thus, adding the residues obtained in equations (\ref{eq-R40}), (\ref{eq-R41}), (\ref{eq-R42}) and (\ref{eq-R43}), we have
\begin{equation}
I_5(a)=  \frac{1}{8 a^2} -\frac{3 \pi}{64a} + \frac{7 \zeta(3)}{8\pi^2}-\frac{5 \pi a}{256} + \cdots.
\end{equation}

\section{Fermionic Pressure Integral \boldmath $I_6(a)$}
\label{sec-F}

In this appendix, we evaluate the fermionic radiation pressure integral (\ref{eq-I6}), expressed as
\begin{equation}
I_6(a)=\int_0^\infty  \ln\left(1+e^{-x}\right) \frac{x^2 }{(a^2+x^2)^3} dx.
\label{eq-I66}
\end{equation}

Defining
\begin{equation}
    f(x)= \ln\left(1+e^{-x}\right) \quad \text{and} \quad g(x)= \frac{x^2 }{\left(a^2+x^2\right)^3},
\end{equation}
their Mellin transforms are expressed as 
\begin{equation}
F(s)=\int_0^\infty x^{s-1}  \ln\left(1+e^{-x}\right)\,dx \quad \text{and} \quad  G(s) = \int_{0}^{\infty} x^{s-1}\,\frac{x^2}{(a^2+x^2)^3}\,dx.
\label{eq-FG6}
\end{equation}

To evaluate $F(s)$ give by (\ref{eq-FG6}), we employ the expansion
\begin{equation}
\ln(1+e^{-x})=\sum_{n=1}^{\infty}\frac{(-1)^{n+1}}{n}e^{-nx},
\end{equation}
so that
\begin{equation}
F(s) = \sum_{n=1}^{\infty}\frac{(-1)^{n+1}}{n} \int_0^\infty x^{s-1}e^{-nx}\,dx.
\end{equation}

Using the Gamma integral $\int_0^\infty x^{s-1}e^{-nx}\,dx=\frac{\Gamma(s)}{n^s}$, we get
\begin{equation}
F(s)= \Gamma(s)\sum_{n=1}^{\infty} \frac{(-1)^{n+1}}{n^{s+1}}  = \Gamma(s)(1-2^{-s})\zeta(s+1),
\label{eq-F6}
\end{equation}
noting that the summation reduces to the Riemann zeta function as before. The Mellin transform $G(s)$ has already been evaluated in equation (\ref{eq-G}).

Using Mellin’s Parseval theorem given by (\ref{eq-convo}), and substituting from equations (\ref{eq-F6}) and (\ref{eq-G}), we find the integral $I_6(a)$ given in (\ref{eq-I66}) through the convolution 
\begin{equation}
I_6(a)  = \frac{1}{2\pi i} \int_{c-i\infty}^{c+i\infty} \Gamma(s)(1-2^{-s})\zeta(s+1) \Gamma\left(\frac{3-s}{2}\right)
\Gamma\left(\frac{3+s}{2}\right)\frac{a^{-(s+3)}}{4}\,ds.
\end{equation}

We choose the contour in the regular region $0 < {\rm Re}\ s <3$ and evaluate this integral using Cauchy's residue theorem. We calculate the residues at the poles $s=0,-1,-2,-3,-4, \cdots$.

{\bf Residue at \boldmath $s=0$.} At $s=0$, there is a double-pole coming from $\Gamma(s)$ and $\zeta(s+1)$. Writing $s=0 +\varepsilon $, the double-pole is revealed by the expansions 
\begin{equation}
\Gamma(0+\varepsilon) = \frac{1}{\varepsilon} - \gamma + \left( \frac{\gamma^2}{2} + \frac{\pi^2}{12} \right) \varepsilon+ \mathcal{O}(\varepsilon^2), \quad \zeta(1+\varepsilon) = \frac{1}{\varepsilon} + \gamma - \gamma_1 \varepsilon + \frac{\gamma_2}{2!} \varepsilon^2 + \mathcal{O}(\varepsilon^3)
\end{equation}
\begin{equation}
1-2^{-\varepsilon} =\varepsilon \ln 2-\frac{\varepsilon^2(\ln 2)^2}{2!}+\mathcal{O}(\varepsilon^3), \quad a^{-(3+\varepsilon)}
=a^{-3}-\varepsilon a^{-3} \ln a+\frac{\varepsilon^2(\ln a)^2}{2!} a^{-3}- \mathcal{O}(\varepsilon^3).
\end{equation}

However, although $(1-2^{-s})=0$ multiples the integrand, there is now a non-zero contribution from this double-pole. The residue at $s=0$ is given by 
\begin{equation}
\mathop{\rm Res}_{s=0}  I_6(a) = \lim_{s\to0}\,
\frac{d}{ds}\left[ (s+0)^2\,
\Gamma(s)(1-2^{-s})\zeta(s+1) \Gamma\left(\frac{3-s}{2}\right)
\Gamma\left(\frac{3+s}{2}\right)\frac{a^{-(s+3)}}{4} \right],
\end{equation}
so that,
\begin{equation}
\mathop{\rm Res}_{s=0}  I_6(a) = \lim_{\varepsilon\to0}\,
\frac{d}{d\varepsilon}\left[ \varepsilon^2\,
\Gamma(\varepsilon)(1-2^{-\varepsilon})\zeta(1+\varepsilon) \Gamma\left(\frac{3-\varepsilon}{2}\right)
\Gamma\left(\frac{3+\varepsilon}{2}\right)\frac{a^{-(\varepsilon+3)}}{4}  \right].
\end{equation}

Since the leading order terms in the expansions contribute to this limit, we get
\begin{align}
\mathop{\rm Res}_{s=0}  I_6(a) =& \lim_{\varepsilon\to0}\,
\frac{d}{d\varepsilon}\left[ \varepsilon^2\,
\left(\frac{1}{\varepsilon} - \gamma \right)\left(\varepsilon \ln 2 \right) \left(\frac{1}{\varepsilon} + \gamma\right) \Gamma\left(\frac{3-\varepsilon}{2}\right)
\Gamma\left(\frac{3+\varepsilon}{2}\right)\frac{(a^{-3}-\varepsilon a^{-3} \ln a)}{4}  \right] \nonumber \\& = \left(\ln 2 \right) \left(\frac{\sqrt{\pi}}{2}\right) \left(\frac{\sqrt{\pi}}{2}\right) \left(\frac{a^{-3}}{4}\right)= \frac{\pi \ln 2}{16a^3}.
\label{eq-R60}
\end{align}

{\bf Residue at \boldmath $s=-1$.} For the pole at $s=-1$, writing $s=-1 +\varepsilon $ reveals the pole of $ \Gamma(s)$ via the expansion $\Gamma(-1+\varepsilon)
=-\frac1{\varepsilon}+\mathcal{O}(1)$. Consequently, the residue at $s=-1$ is given by 
\begin{align}
 \operatorname*{Res}_{s=-1}I_6(a)&=  \lim_{\varepsilon \to 0}  \left[ \varepsilon \left(-\frac{1}{\varepsilon} \right) (1-2^{1-\varepsilon})\zeta(\varepsilon) \Gamma\left(\frac{4-\varepsilon}{2}\right) \Gamma\left(\frac{2+\varepsilon}{2}\right) \frac{a^{-(2+\varepsilon)}}{4} \right] \nonumber \\&
 =\left(-1 \right)  \left(-1 \right) \left(-\frac{1}{2}\right) \left(1 \right) \left( 1 \right) \frac{a^{-2}}{4}=-\frac{1}{8 a^2}.
\label{eq-R61} 
\end{align}

{\bf Residue at \boldmath $s=-2$.} For the pole at $s=-2$, writing $s=-2 +\varepsilon $ reveals the pole of $ \Gamma(s)$ via the expansion $\Gamma(-2+\varepsilon)
=\frac1{2\varepsilon}+\mathcal{O}(1)$. Consequently, the residue at $s=-2$ is given by 
\begin{align}
 \operatorname*{Res}_{s=-2}I_6(a)&=  \lim_{\varepsilon \to 0}  \left[ \varepsilon \left(\frac{1}{2\varepsilon} \right) (1-2^{2-\varepsilon})\zeta(-1+\varepsilon) \Gamma\left(\frac{5-\varepsilon}{2}\right) \Gamma\left(\frac{1+\varepsilon}{2}\right) \frac{a^{-(1+\varepsilon)}}{4} \right] \nonumber \\&
 =\left(\frac{1}{2} \right)  \left(-3 \right) \left(-\frac{1}{12}\right) \left(\frac{3\sqrt{\pi}}{4} \right) \left( \sqrt{\pi} \right) \frac{a^{-1}}{4}=\frac{3 \pi}{128 a}.
\label{eq-R62} 
\end{align}

{\bf Residue at \boldmath $s=-3$.} At $s=-3$, there is a double-pole coming from $\Gamma(s)$ and $\Gamma\left(\frac{3+s}{2}\right)$. Writing $s=-3+\varepsilon $, the double-pole is revealed by the expansions 
\begin{equation}
\Gamma(-3+\varepsilon) = -\frac{1}{6\varepsilon} + A  + B\varepsilon + \mathcal{O}(\varepsilon^2), \quad
\zeta(-2+\varepsilon)  = \varepsilon\zeta'(-2)+\mathcal{O}(\varepsilon^2),
\end{equation}
\begin{equation}
\Gamma\left(3 - \frac{\varepsilon}{2}\right) = 2 - \left(\frac{3}{2} - \gamma\right) \varepsilon + \mathcal{O}(\varepsilon^2), \quad
\Gamma\left(\frac{\varepsilon}{2}\right) = \frac{2}{\varepsilon} - \gamma + C \varepsilon + \mathcal{O}(\varepsilon^2),
\end{equation}
and
\begin{equation}
a^{-\varepsilon}=e^{-\varepsilon\ln a}=1-\varepsilon\ln a + \varepsilon^2\ln^2 a + \mathcal{O}(\varepsilon^3),
\end{equation}
where
\begin{equation}
A=\left( \frac{\gamma}{6} - \frac{11}{36} \right), \quad
B=- \left( \frac{\gamma^2}{12} - \frac{11\gamma}{36} + \frac{85}{216} + \frac{\pi^2}{72} \right), \quad C=\frac{1}{4}\left(\gamma^2 + \frac{\pi^2}{3}\right).
\end{equation}

However, although $\zeta{(-2)}=0$ multiples the integrand, there is now a non-zero contribution from this double-pole. The residue at $s=-3$ is given by 
\begin{equation}
\mathop{\rm Res}_{s=-3}  I_6(a) = \lim_{s\to0}\,
\frac{d}{ds}\left[ (s+3)^2\,
\Gamma(s)(1-2^{-s})\zeta(s+1) \Gamma\left(\frac{3-s}{2}\right)
\Gamma\left(\frac{3+s}{2}\right)\frac{a^{-(s+3)}}{4} \right],
\end{equation}
so that,
\begin{equation}
\mathop{\rm Res}_{s=0}  I_6(a) = \lim_{\varepsilon\to0}\,
\frac{d}{d\varepsilon}\left[ \varepsilon^2\,
\Gamma(-3+\varepsilon)(1-2^{(3-\varepsilon)})\zeta(-2+\varepsilon) \Gamma\left(\frac{6-\varepsilon}{2}\right)
\Gamma\left(\frac{\varepsilon}{2}\right)
\frac{a^{-\varepsilon}}{4}  \right].
\end{equation}

Since the leading order terms in the expansions contribute to this limit,
\begin{align}
\mathop{\rm Res}_{s=-3}  I_6(a) =& \lim_{\varepsilon\to0}\,
\frac{d}{d\varepsilon}\left[ \varepsilon^2\,
\left(-\frac{1}{6\varepsilon} \right)\left(-7 \right) \left\{\varepsilon\zeta'(-2)\right\} \left(2\right)
\left(\frac{2}{\varepsilon}\right)\frac{(1)}{4}  \right] =- \frac{7\zeta(3)}{24\pi^2},
\label{eq-R63}
\end{align}
using $\zeta'(-2)=-\frac{\zeta(3)}{4 \pi^2}$.

{\bf Residue at \boldmath $s=-4$.} For the pole at $s=-4$, writing $s=-4 +\varepsilon $ reveals the pole of $ \Gamma(s)$ via the expansion $\Gamma(-4+\varepsilon)
=\frac1{24\varepsilon}+\mathcal{O}(1)$. Consequently, the residue at $s=-4$ is given by 
\begin{align}
 \operatorname*{Res}_{s=-4}I_6(a)&=  \lim_{\varepsilon \to 0}  \left[ \varepsilon \left(\frac{1}{24\varepsilon} \right) (1-2^{4-\varepsilon})\zeta(-3+\varepsilon) \Gamma\left(\frac{7-\varepsilon}{2}\right) \Gamma\left(\frac{-1+\varepsilon}{2}\right) \frac{a^{(1-\varepsilon)}}{4} \right] \nonumber \\&
 =\left(\frac{1}{24} \right)  \left(-15 \right) \left(\frac{1}{120}\right) \left(\frac{15\sqrt{\pi}}{8} \right) \left(-2 \sqrt{\pi} \right) \frac{a}{4}=\frac{5 \pi a}{1024}.
\label{eq-R64} 
\end{align}

{\bf Total Contribution to \boldmath $I_6(a)$.} Adding the residues obtained in equations (\ref{eq-R60}), (\ref{eq-R61}), (\ref{eq-R62}), (\ref{eq-R63})and (\ref{eq-R64}), we have 
\begin{equation}
I_6(a)= \frac{\pi \ln 2}{16a^3}-\frac{1}{8a^2}+\frac{3\pi}{128 a} - \frac{7\zeta(3)}{24\pi^2} + \frac{5 \pi a}{1024} + \cdots.
\end{equation}


\end{document}